\documentclass[prx,notitlepage,twocolumn,superscriptaddress]{revtex4-1}

\usepackage{epsfig,color}
\usepackage{graphicx}
\usepackage{dcolumn}
\usepackage{bm}
\usepackage{amsmath, amsfonts, amssymb,mathrsfs}
\usepackage{pstricks}
\usepackage{amsxtra}
\usepackage{amsthm}
\usepackage{natbib}
\usepackage{qcircuit}
\usepackage{array}

\newcommand{\ket}[1]{\mbox{$|#1\rangle$}}
\newcommand{\bra}[1]{\mbox{$\langle#1|$}}

\def\be{\begin{equation}}      
\def\ee{\end{equation}}
\def\beu{\begin{equation*}}   
\def\eeu{\end{equation*}}

\DeclareMathOperator{\trace}{Tr}      

\providecommand{\mean}[1]{\langle#1\rangle}

\newtheorem{theorem}{Theorem}[section]
\theoremstyle{definition}

\definecolor{new}{rgb}{.08,.05,.8}

\newcommand{\delete}[1]{{}}

\begin{document}
\title{Dynamical purification phase transitions induced by quantum measurements}
\date{\today}
\author{Michael J. Gullans}
\affiliation{Department of Physics, Princeton University, Princeton, New Jersey 08544, USA}
\author{David A. Huse}
\affiliation{Department of Physics, Princeton University, Princeton, New Jersey 08544, USA}
\affiliation{Institute for Advanced Study, Princeton, New Jersey 08540, USA}

\begin{abstract}
Continuously monitoring the environment of a quantum many-body system reduces the entropy of (purifies) the reduced density matrix of the system, conditional on the outcomes of the measurements.  We show that, for mixed initial states, a balanced competition between measurements and entangling interactions within the system can result in a dynamical purification phase transition between (i) a phase that locally purifies at a constant system-size-independent rate, and (ii) a ``mixed'' phase where the purification time diverges exponentially in the system size.    The residual entropy density in the mixed phase implies the existence of a quantum error-protected subspace where quantum information is reliably encoded against the future non-unitary evolution of the system.  We show that these codes are of potential relevance to fault-tolerant quantum computation as they are often highly degenerate and satisfy optimal tradeoffs between encoded information densities and error thresholds.  In spatially local models in $1+1$ dimensions, this phase transition for mixed initial states occurs concurrently with a recently identified class of entanglement phase transitions for pure initial states.  The purification transition studied here also generalizes to systems with long-range interactions, where conventional notions of entanglement transitions have to be reformulated.  We numerically explore this transition for monitored random quantum circuits in 1+1 dimensions and all-to-all models.  Unlike for pure initial states, the mutual information of an initially completely-mixed state in 1+1 dimensions grows sublinearly in time due to the formation of the error protected subspace.   Purification dynamics is likely a more robust probe of the transition in experiments, where imperfections generically reduce entanglement and drive the system towards mixed states.  We describe the motivations for studying this novel class of non-equilibrium quantum dynamics in the context of advanced quantum computing platforms and fault-tolerant quantum computation.
\end{abstract}
\maketitle


\section{Introduction}

In thermodynamic equilibrium, pure quantum states can only be achieved at absolute zero temperature.  The nonequilibrium thermodynamic cost of purification is encoded in the third law of thermodynamics, which states that it is impossible to reach a zero entropy (pure) quantum state in a finite amount of time.  In quantum information science, purification plays an essential role in many models of quantum computation, where one often assumes access to highly pure computational or ancilla qubits \cite{NielsenChuang}.  Although it is known that the requisite purification is possible given sufficiently fine control over a quantum system and its environment \cite{Bennett96,Cirac99,Schulman99,Bravyi05}, the question of whether a generic interacting many-body quantum system coupled to a finite temperature bath (i.e., an open quantum system) can be driven to a pure state remains less understood \cite{Ticozzi14,Masanes17}.

An essential resource in controlling open quantum systems is the ability to make measurements of the system, which can then be used to perform feedback and conditional control (e.g., the famous Maxwell demon) \cite{WisemanBook}.  Purification, however, does not require any feedback because the continuous monitoring of the environment can be used to continually gain information about the system; thereby, reducing the number of accessible states consistent with the measurement record and the intermediate dynamics  \cite{Zoller87,CarmichaelBook,Plenio98}.  Naively, one expects that  continuous, perfect monitoring will rapidly purify the system; however, it is known from the study of quantum error correcting codes that quantum states can be protected from extensive numbers of local measurements \cite{Calderbank96,Steane96,Preskill98}.  Recently, there has been significant experimental progress towards realizing the requisite ingredients for such measurement-driven purification of many-body states in quantum computing platforms \cite{Hume07,Negnevitsky18,Jiang09,Neumann10,Vijay11,Lange14,Ofek16,Minev18,Mi18,Nakajima19}. 

In this Article, we show that there is a dynamical purification phase transition as one changes the measurement rate in a class of random quantum circuit models with measurements.  For pure initial states, it was  shown recently that there is an entanglement phase transition in these models from area-law to volume-law entanglement  \cite{Li18,Skinner18,Chan18b}, with subsequent work deepening our understanding of this family of entanglement phase transitions \cite{Li19,Szyniszewski19,Choi19}.  We show that, for mixed initial states, this entanglement phase transition transforms into a dynamical purification transition between a ``pure'' phase, with a constant purification rate in the thermodynamic limit, and a ``mixed'' phase, where the purification time diverges exponentially in system size.  Thus, if one takes the simultaneous limit of an infinite system and infinite time, with any power-law relation between system size and time, then an initially maximally-mixed state has a nonzero long-time entropy density in the mixed phase, while it becomes pure, and area-law entangled, in the pure phase.   

We provide a more general definition of purification transitions in terms of a phase transition in the quantum channel capacity of the underlying open system dynamics.  This definition can be applied to arbitrary quantum channels and implies that a purification transition should be  fundamentally interpreted as a type of quantum error correction threshold.  Our results, therefore, help further establish the connections between measurement-induced phase transitions, channel capacities, and quantum error correction \cite{Choi19}.   We additionally strengthen these connections 
by showing that the unitary-measurement dynamics projects the system into an optimal quantum error correcting code space.  This encoding achieves the capacity for the future evolution of the channel in single-use error correction.  Thus, our work points to a large family of previously unexplored codes with an optimal tradeoff between  code rates and error thresholds, with relatively simple encoding schemes, that may provide useful insight or applications to fault-tolerant quantum computation. 

To develop the basic phenomenology of purification phase transitions, we numerically explore the measurement-induced transition in the $1+1$-dimensional stabilizer circuit model of Ref.~\cite{Li19}.  We find that a tripartite mutual information, or topological entanglement entropy \cite{Kitaev06,Levin06}, allows a scaling analysis with substantially reduced finite-size effects compared to other metrics, allowing more precise estimates of the critical behavior.  Unlike the rapid linear-in-time growth of entanglement that can occur for pure initial states, the bipartite mutual information of maximally-mixed initial states grows sublinearly in time.  This slow growth of the bipartite mutual information arises from the natural formation of the quantum error corrected code space described above, which protects the system from measurement-induced collapse of coherent quantum information.  

The dynamics of quantum information in systems with  long-range interactions is currently an active area of research  \cite{Hastings06,Richerme14,Fossfeig15,Zhou19b,Chen19,Tran19,Else20}.   The extreme limit of  all-to-all coupled systems have no notion of spatial locality or a well-defined geometry, which requires one to revisit the conventional definition of an area-to-volume-law entanglement transition.  Interestingly, however, the purification transition we find naturally persists in all-to-all coupled models with $k$-local interactions. 

Due to the difficulty in isolating and measuring volume-law entanglement, purification dynamics  can also serve as  a more robust probe of measurement-induced entanglement transitions in experiments.  For example, we recently found that the purification dynamics of a finite number of qubits acts as a local order parameter \cite{Gullans19e}, which, together with the quantum Fisher information \cite{Bao19}, allows direct experimental access to the phase transition on near-term quantum computing devices.
We argue that an important motivation for studying this class of non-equilibrium quantum dynamics is to  develop more efficient routes to fault-tolerant quantum computation.


\section{Overview}
\label{sec:overview}

Before describing our  analysis, we present an overview of the main results on purification transitions and dynamically generated codes.  Our definition of purification phase transitions in open system dynamics fundamentally relies on a quantity known as the quantum channel capacity, which determines the maximum amount of quantum information that can be transmitted by a noisy quantum channel  \cite{Lloyd97,HolevoBook,Devetak05,Devetak05b}.  A closely related quantity is the coherent quantum information, which plays an important role in the basic theory of quantum error correction \cite{Schumacher96,Barnum02,Schumacher01}.  

\begin{figure}[tb]
\begin{center}
\includegraphics[width = .49 \textwidth]{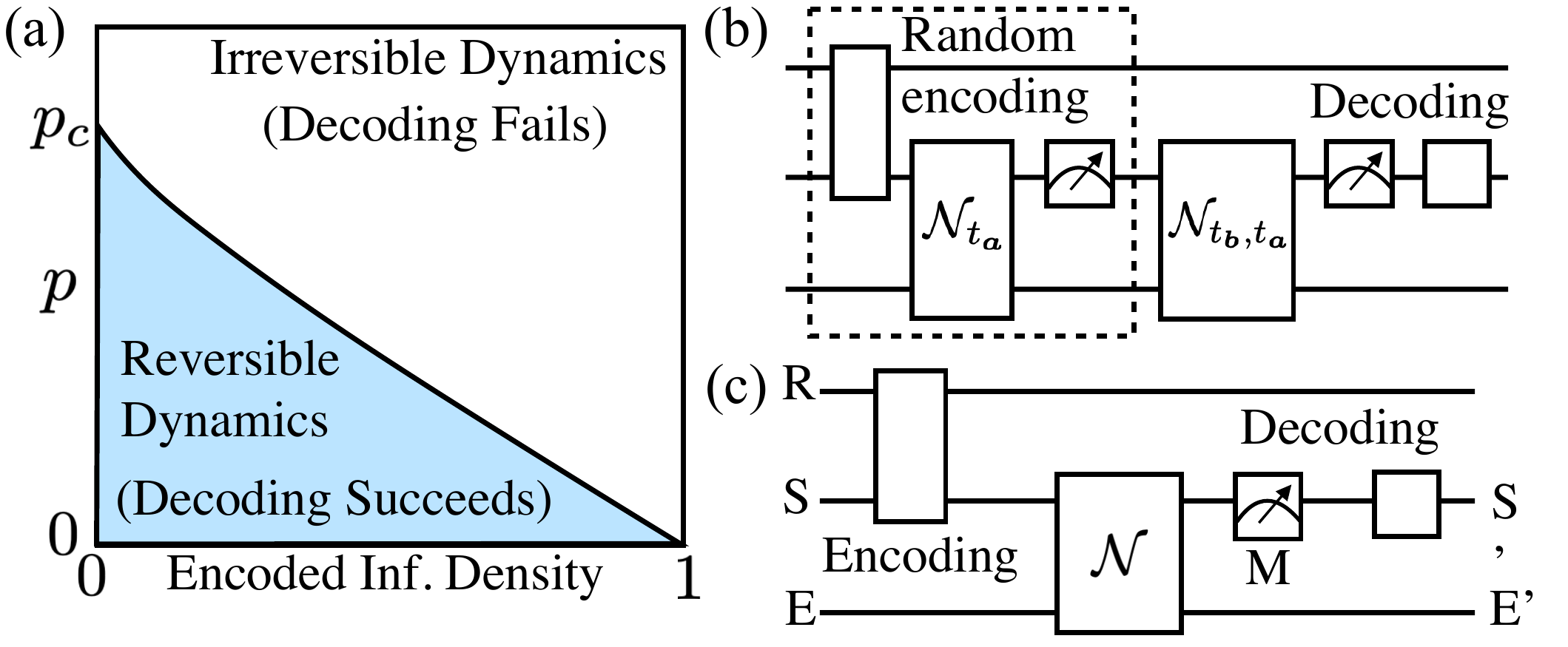}
\caption{(a)  Qualitative phase diagram for a purification phase transition.  Inside the blue region it is possible to encode an extensive amount of quantum information and recover the initial state for all polynomial times in the thermodynamic limit with fidelity arbitrarily close to one.  Outside this boundary there is no encoding and decoding pair that succeeds and the dynamics is fundamentally irreversible. (b) Basic setup of the encoding/decoding problem for the class of channels studied in this work where the capacity-achieving encoding is implemented by the dynamics of the channel itself. (c) General model for open system dynamics using a reference $R$ system and an environment $E$ to purify a quantum channel $\mathcal{N}$.  In the language of quantum error correction, the initial state preparation is the encoding operation, while subsequent measurements and control are the decoding operation.   }
\label{fig:channel}
\end{center}
\end{figure}

 In Fig.~\ref{fig:channel}(a), we provide a qualitative phase diagram for a purification transition as a function of a generalized measurement rate $p$.  For $p>p_c$, there is no pair of encoding and decoding operations that can protect an extensive amount of quantum information. Thus, the system always forgets initial conditions (purifies) and the dynamics is fundamentally irreversible.  On the other hand, for $p<p_c$, there are extensive subspaces that can be arbitrarily well protected on all polynomial timescales in the thermodynamic limit through quantum error correction, i.e., the system  remembers initial conditions.  The maximum amount of quantum information that can be stored in the system (depicted by the boundary line in the figure) is given by the channel capacity.

The connection between measurement-induced phase transitions, quantum channel capacities, and quantum error correction  was previously explored by Choi, Bao, Qi and Altman  using a simplified model with either a single round of measurements at the end of the circuit or nearly fully scrambling dynamics on large blocks of qubits between each round of measurements \cite{Choi19}.  We show how these connections can be established in a more general context by considering the mixed-state dynamics of the underlying quantum channel.

Furthermore, we rigorously prove a sufficient set of  conditions under which the monitored channel dynamics can act as an optimal encoding for the future evolution of the system.  We show numerically that these conditions are satisfied for a representative family of models that exhibit measurement-induced entanglement transitions. The associated high-fidelity recovery operations apply to a single use of the channel.  Thus, the system in the mixed phase generates a family of quantum error correcting codes that saturate fundamental bounds on the tradeoff between the density of encoded quantum information and the associated error threshold. 

The basic concept underlying our theorem is illustrated in Fig.~\ref{fig:channel}(b).   We first run a \textit{monitored channel} (defined below) for a particular mixed state input $\rho_0$ on a time scale polynomial in the system size.  The repeated rounds of measurements, followed by unitary scrambling dynamics, randomly projects the system into a quantum error correcting code space density matrix $\rho_m$.  Remarkably, this encoding process is successful for the future evolution of the channel that is statistically independent from the past evolution.   Stated more precisely, a random ``codeword'' $\ket{c_i}$ drawn from any ensemble representation of $\rho_m = \sum_i \lambda_i \ket{c_i}\bra{c_i}$ can be approximately recovered back to this state following further evolution with the channel. For a correct choice of $\rho_0$, this encoding scheme protects the maximal possible amount of quantum information in the thermodynamic limit.  These results have broad implications for the study of measurement-induced phase transitions, as we describe in this Article.

The classification of the complete family of information theoretic phases of unitary-measurement dynamics is a fascinating and rich problem, with potential applications to fault-tolerant quantum computation.  In Sec.~\ref{sec:disc}, we discuss a variety of motivations to better characterize these phases and the associated quantum error correcting codes in the ordered phase, as well as investigate these dynamics in intermediate-scale quantum devices.  After reading Sec.~\ref{sec:purintro}A-B, those readers  less interested in the proofs and further details of our analysis can avoid the intervening sections and Appendixes, or return to them later.

The paper is organized as follows:    In Sec.~\ref{sec:purintro}, we introduce the random circuit model studied in this work and establish the basic phenomenology of purification transitions in unitary-measurement dynamics.  We then provide a more general definition of a purification transition in terms of the long-time scaling of the quantum channel capacity with system parameters in the thermodynamic limit. To make more explicit connections  to unitary-measurement models, we introduce the notion of strong purification transitions, unitary-dephasing channels, and monitored channels.  In Sec.~\ref{sec:codethm}, we show rigorously that, for monitored channels with a strong purification transition, the late time density matrix defines a quantum error correcting code space that can saturate the channel capacity bound for the future evolution of the system, with potentially useful properties for fault-tolerant quantum computation. Furthermore, we show that the dynamics in the ordered phase has interesting parallels to more conventional decoding problems in quantum error correction.  In Sec.~\ref{sec:traj}, we discuss the class of monitored random circuit models that have been broadly found to exhibit measurement-induced entanglement transitions, which we study in this work.  We also describe the relation between between quantum trajectories and our definition of a purification transition.  In Sec.~\ref{sec:crit}, we present a detailed overview of the critical properties of the entanglement and purification transition in $1+1$ dimensions based on a finite-size scaling analysis of numerical simulations. 

In Sec.~\ref{sec:bob}, we introduce a family of two-local, infinite range models that display clear signatures of the purification transition. We find that the transition in this model violates a version of the Hamming bound on $p_c$ \cite{Fan20}, indicating that the zero-rate codes near the critical point are highly degenerate.  Degenerate, zero-rate quantum codes, such as the surface code \cite{Dennis02}, can have high error correction thresholds and  resilience to errors in gates and measurements, making them potentially useful for fault-tolerant quantum computation.
In Sec.~\ref{sec:disc}, we discuss the general phase diagram and universality classes of measurement-induced transitions opened up by the purification perspective presented here.  We then discuss the implications of our work for fault-tolerant quantum computation and experimental studies of measurement-induced phases in intermediate-scale quantum devices.  We present our conclusions in Sec.~\ref{sec:conc}.  

In Appendix \ref{app:proof1}, we present a formal statement and proof of our theorem on dynamically generated codes.  In Appendix \ref{app:Ic}, we derive a bound on the channel capacity by the average entropy of  unravelings of the channel into quantum trajectories.  In Appendixes \ref{app:stab1}-\ref{app:ent}, we provide more details on the stabilizer formalism, mixed stabilizer states, and methods to compute entropies of stabilizer states.  In Appendix \ref{app:codelength}, we introduce a quantity that can be useful in  characterizing the optimal codes generated by the dynamics that we define as the contiguous code length.    

\section{Purification Transitions}
\label{sec:purintro}
 
In this section, we first present an overview of the phenomenology of purification transitions in the random circuit models studied in this work, which are based on stabilizer circuit dynamics.  Such models were introduced in the context of measurement-induced entanglement transitions by Li, Chen, and Fisher \cite{Li18,Li19}.  We  then provide a general definition of purification transitions in terms of the scaling of the channel capacity in the thermodynamic limit.  This definition is  essentially a reformulation of a quantum error correction threshold, but introducing it in this setting allows us to formulate the necessary concepts and terminology starting from basic concepts in open quantum systems.  To make more concrete connections to unitary-measurement models, we introduce the notion of strong purification transitions, unitary-dephasing channels, and monitored channels.  Such channels satisfy a weaker version of the Knill-Laflamme conditions for quantum error correction \cite{NielsenChuang,Knill00} and form the basis for our theorem on dynamically generated quantum error correcting codes.

\subsection{Random Clifford Model}
\label{sec:stab}

Measurement-induced phase transitions in ensembles of quantum trajectories are now understood to generically appear whenever there is a balanced competition between unitary dynamics that grows entanglement of the system and a measurement process that reduces the entanglement.  To establish the connection to purification transitions we, therefore, study the ``random Clifford'' model introduced in Ref.~\cite{Li19} [see Fig.~\ref{fig:model}(a)], where the properties of the entanglement transition have been most well established due to the ability to perform large-scale numerics.  

The model consists of a ``brickwork'' circuit of random two-site unitaries drawn uniformly from the Clifford group, operating on a linear chain of $L$ qubits with periodic boundary conditions.  In between each layer of unitaries, each site is measured in the $Z$ basis with a fixed probability $p$.  One tunes through the phase transition by changing $p$.  
Including the measurements, this random circuit is an example of a stabilizer circuit, which starting from the computational zero state or any stabilizer state, can be simulated on a classical computer in a time that scales polynomially in $L$ \cite{Gottesman98,Aaronson04}.  As a result, one can perform a finite-size scaling analysis of the transition for hundreds or even thousands of qubits.  Furthermore, polynomial-time classical algorithms have been introduced to compute  entropies and mutual information of stabilizer states \cite{Audenaert05,Nahum16}.  
In Appendix \ref{app:stab1}-\ref{app:ent}, we provide a more detailed overview of the formalism used to describe stabilizer circuits and states. 

Although these special properties of stabilizer circuits make many aspects of their dynamics nongeneric, one of the defining features of the Clifford group is that it forms a unitary $t$-design for $t\le 3$ \cite{DiVincenzo02,Webb16}.  This property implies that channel-averaged properties of Clifford models often have similar phenomenology to more general quantum chaotic models \cite{Nahum16,Nahum17,vonKeyserlingk17}.  The initial state can either be a mixed or pure stabilizer state, where a mixed stabilizer state is defined as the uniform mixture of all pure stabilizer states associated to a given stabilizer group (see Appendix \ref{app:mixedevolve}).  A mixed stabilizer state with rank $2^{k}$ has an equivalent interpretation as a projector onto an $[N,k]$ stabilizer code, where $k$ refers to the number of logical, or encoded, qubits in the code.

\subsection{Purification Phases}
\label{sec:phases}

\begin{figure}[tb]
\begin{center}
\includegraphics[width = .4 \textwidth]{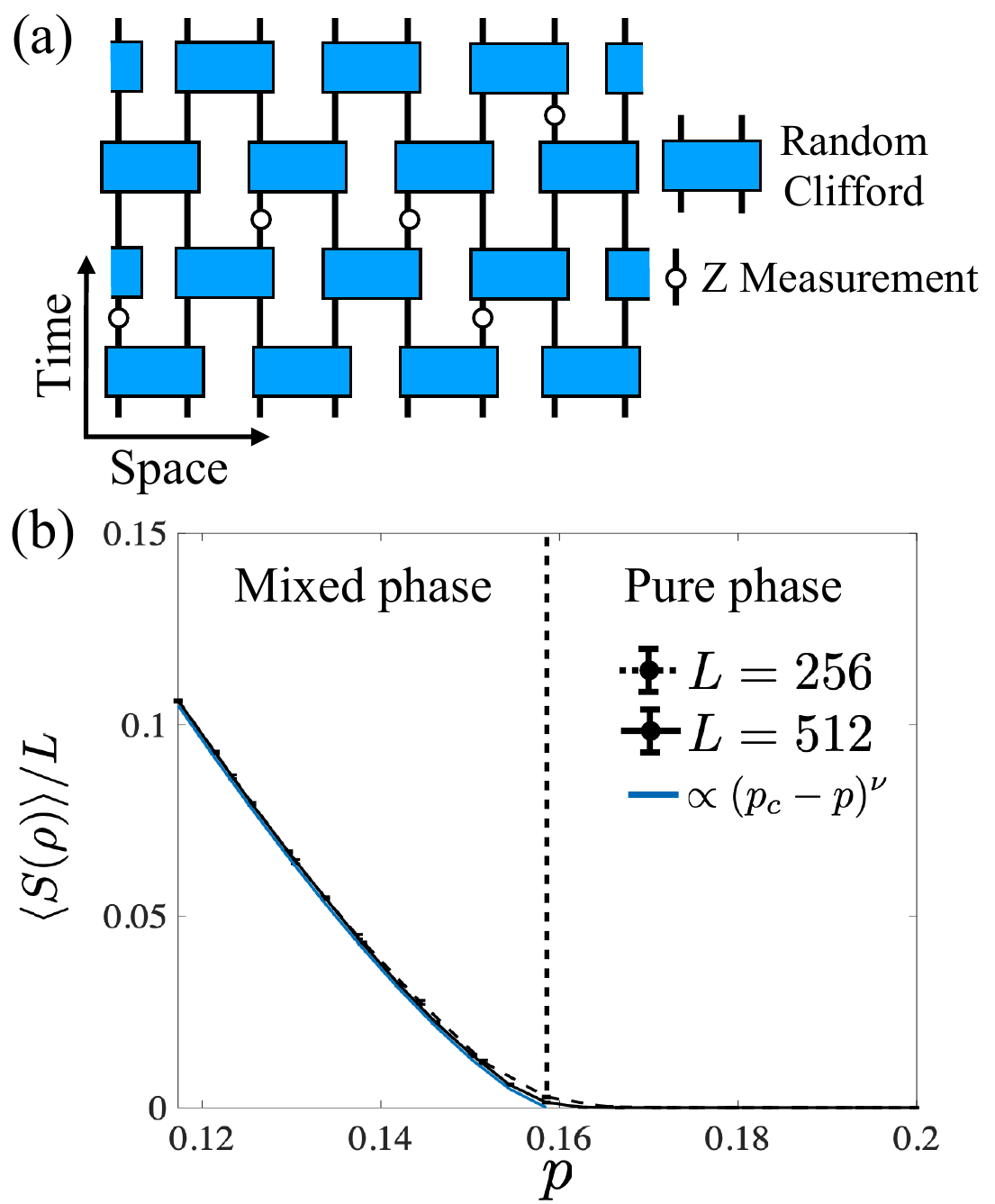}
\caption{(a) Random quantum circuit model studied in this work called the ``random Clifford'' model \cite{Li18}.  Local two-qubit unitaries are drawn uniformly from the Clifford group in a 1D brickwork arrangement with periodic boundary conditions.   
Between each layer of gates, projective $Z$ measurements happen with probability $p$ at each site.  (b) Phase diagram for the late-time, circuit-averaged entropy density $\mean{S(\rho)}/L$ starting from the completely-mixed initial state.  We took time $t=4L$ and $L=256$ and 512 to limit finite-size effects, although signatures of the transition appear already at $L=8$.  Blue curve shows $A (p_c - p)^\nu$ for $A \approx 7.3$, $p \le p_c = 0.1593(5)$ and $\nu = 1.28(2)$ obtained below. }
\label{fig:model}
\end{center}
\end{figure}

We now describe the basic signatures of the phases in the purification transition.  One key signature of the  transition is shown in Fig.~\ref{fig:model}(b).  Here, we take $L=256$ and 512 sites and, starting from the completely-mixed initial state, $\rho = \mathbb{I}/2^L$, we run many realizations of the random circuit out to a time $t$ (= number of two-site unitaries that have acted on each qubit)
that is a fixed multiple of $L$.  We then compute the entropy density of the resulting state $\langle S(\rho)\rangle/L$ averaged over random circuit realizations, assuming perfect knowledge of the outcome of all measurements in each run of the circuit \cite{Li18,Skinner18,Chan18b}.  For each such stabilizer circuit started from this completely-mixed initial state, the von Neumann and all Renyi entropies are equal at a given time, although they differ between circuits and can decrease with time.  Below a critical value of $ p = p_c = 0.1593(5)$ (determined to this level of precision below), we see that the late time density matrix has residual entropy density that is independent of $L$ with the scaling $\langle S(\rho)\rangle/L \sim (p_c- p)^\nu$ for $\nu = 1.28(2)$.  
In Sec.~\ref{sec:traj}, we show that this residual entropy density directly implies an extensive channel capacity (see Sec.~\ref{sec:pur} for a definition).  
Although we cannot run the dynamics for these sizes out to exponentially long times, our finite-size scaling analysis in Sec.~\ref{sec:crit} is consistent with an exponentially divergent lifetime of the plateau value of $S(\rho)$.

On the other hand, for $p> p_c$, the average entropy density decays to zero with a decay rate that is independent of $L$ (leading to a $\sim \ln{L}$ purification time). To the level of precision we can test, the values of $p_c$ and $\nu$ for the purification transition are identical to those for the entanglement phase transition for pure initial states.   In Sec.~\ref{sec:crit}, we provide a detailed overview of the properties of the system near the critical point.  We now give some intuition for the physical origin of the two purification phases deep in their respective regimes.

The basic origin of the pure phase can be simply understood for $p$ sufficiently close to one.  In this limit, each layer of measurements  projects the system into a near perfect product state in the $Z$ basis. As a result, any correlations and complexity in the system can  build up only over a few sites before being decohered by the measurements, which makes the system highly insensitive to initial conditions.    Since pure states are a fixed manifold of the dynamics, the system will rapidly converge to zero entropy density, regardless of initial conditions.  

The mixed phase generally has a richer many-body dynamics than the pure phase.  It turns out some basic features of the mixed phase already appear when the measurements occur only at a single given site at an arbitrarily slow rate (technically, a sufficient condition for large $L$ is $p \ll 1/L^3$ \cite{Brandao16}).  In this limit, the spatial structure of the circuit is largely irrelevant for the late time dynamics, and we can replace the unitary between measurements by a random Clifford gate that acts on the entire set of $L$ qubits.  Starting from the completely-mixed initial state, the density matrix after the first measurement is
\be
\rho_1 = \frac{1}{2^L}(\mathbb{I} + m_1 Z), 
\ee
where 
$m_1 = \pm 1$ is the first measurement outcome , and $Z$ is the Pauli-$Z$ matrix on the site measured.  The purity of the system has increased by a factor of 2.  Following the intermediate time dynamics and second measurement, the density matrix is updated as
 {\small
 \be \label{eqn:rho2}
\rho_2 = \frac{P_2 U_2 \rho_1 U_2^\dag P_2 }{\trace[U_2^\dag P_2 U_2 \rho_1] } = \frac{1}{2^L} (\mathbb{I} + m_2 Z + 2 m_1 P_2 U_2 Z U_2^\dag P_2),
\ee
}\noindent
where $P_n = \frac{1}{2}(\mathbb{I} + m_n Z)$ is a projector onto the state consistent with the outcome of measurement $n$ of $m_n =\pm 1$ and $U_n$ is the unitary for the random circuit between measurement $n-1$ and $n$.  To compute the denominator of Eq.~(\ref{eqn:rho2}), we assume that $U_2 Z U_2^\dag \ne Z$ (this is true with probability $1 - \frac{1}{4^{L}-1}$ since $U_2$ maps $Z$ to a random traceless product of Pauli operators on all $L$ qubits),  such that $P_2 U_2 Z U_2^\dag$ has zero trace.  Using the property that the Clifford group is a 2-design, we can compute the circuit averaged purity after this second measurement
\be
\langle \trace[\rho_2^2] \rangle = 3 /2^L.
\ee
Extending these arguments to many subsequent measurements, one finds that, each time a measurement occurs, it increases the average purity by only  $1/2^L$ so
\be
\langle \trace[\rho_n^2] \rangle = (n+1)/2^L, ~ n \ll 2^L~,
\ee
leading to an exponentially long purification time.  
The average entropy satisfies the inequality
\be
\mean{S(\rho_n)} \ge -\log \langle \trace[\rho_n^2] \rangle = L - \log (n+1).
\ee
This limiting case establishes some essential features of the measurement-induced dynamics in the mixed phase, including the insensitivity of the basic phenomenology to spatial locality.  
Although some aspects of the above argument extend to small but $L$-independent $p$, to establish the existence of the mixed phase in the present work, we rely on  numerical solutions of the model.  In addition to 1+1-dimensional models, we also study the mixed phase in all-to-all models in Sec.~\ref{sec:bob}.

\subsection{Purification Transitions: Definitions}
\label{sec:pur}

In this section, we present a mathematical definition of purification transitions, including a refinement to strong purification transitions, unitary-dephasing channels, and monitored channels that describe the models that have been found to exhibit measurement-induced entanglement transitions. The condition for a unitary-dephasing channel is shown to be equivalent to a weaker version of the Knill-Laflamme conditions for quantum error correction.

In the most general formulation of the problem, we are considering an open, quantum many-body system undergoing time evolution of the form
\begin{align} \label{eqn:Ntgen}
\mathcal{N}_t(\rho) &=T_t \circ \cdots T_1 (\rho),
\end{align}
where $T_i$ are quantum channels or completely-positive-trace-preserving (CPTP) maps \cite{NielsenChuang}.  The models we study below are random and Markovian in the sense that the $T_i$ are generated from independent, identically distributed ensembles, but we do not rely on this property in developing the concept of a purification transition.

 As illustrated in Fig.~\ref{fig:channel}(c), we can always consider a purification of a quantum channel $\mathcal{N}$ by using a three component system consisting of a reference $R$, system $S$, and an environment $E$.  Within this framework, the initial mixed state of the system $\rho = \sum_k \lambda_k \ket{k_S} \bra{k_S}$ is prepared through an  entanglement operation between $R$ and $S$, with $E$ in the computational zero state
\be
\ket{\psi_{RSE} } =\Big( \sum_k \sqrt{\lambda_k}\, \ket{k_R} \otimes \ket{k_S}\Big) \otimes \ket{0}.
\ee
The open system dynamics is  now modeled by an isometry on $SE$ $U_{\mathcal{N}}$ called an isometric embedding 
\be
\ket{\psi_{RS^{'}E^{'}}} = \sum_k \sqrt{\lambda_k}\, (\mathbb{I}_R \otimes U_{\mathcal{N}})  \ket{k_R} \otimes \ket{k_S} \otimes \ket{0},
\ee
where $\mathcal{N}_t(\rho) = \trace_{RE'}[\ket{\psi_{RS^{'}E^{'}}}\bra{\psi_{RS^{'}E^{'}}}]$ and the primes indicate that $U_{\mathcal{N}}$ has been applied to $S$ and $E$.
The original quantum channel dynamics is recovered by tracing over $R$ and $E'$.  
For simplicity, we focus on qubit models in which $R$, $S/S'$, and $E/E'$ consist of tensor-product Hilbert spaces of two-level systems.

A crucial result in quantum information theory is that, when the mutual information  between the reference and the environment is zero, then there exists a perfect recovery operation acting only on the system that can recover the initial entangled state $\ket{\psi_{RSE}}$ \cite{Schumacher96}.  To review the argument, note that zero mutual information between $R$ and $E'$ implies that their reduced density matrix factorizes
\be
\rho_{RE'} = \rho_R \otimes \rho_{E'},
\ee
where $\rho_A = \trace_{A^c} [ \rho]$ is the reduced density matrix on $A$ and $A^c$ is the complement of $A$.  This factorization implies the existence of a Schmidt decomposition for the pure state of the form
\be
\ket{\psi_{RS'E'}} = \sum_{k,\ell } \sqrt{\lambda_k p_\ell} \ket{k_R} \otimes \ket{\psi_{k \ell }} \otimes \ket{\ell }
\ee
where $\ket{k_R}$, $\ket{\psi_{k \ell}}$, and $\ket{\ell}$ are all orthonormal states.  By performing projective measurements of the operators
\be \label{eqn:Ml}
M_{\ell} = \sum_k \ket{\psi_{kl}} \bra{\psi_{kl}}
\ee
and applying the unitary operation depending on the measurement outcome $U_{\ell} \ket{\psi_{kl}} = \ket{k_S}$, we can completely recover the initial state.   The recovery operation acts only on $S$ and can correct any initial state in the support of $\rho$.  As we discuss later, such arguments can be  generalized to the more physically relevant situation of approximate quantum error correction, where $\rho_{RE'}$ only approximately factorizes \cite{Barnum02,Schumacher01}.  

To quantify the notion of recoverability or reversibility of the system in more general settings, we introduce the single-use quantum channel capacity defined in terms of the  coherent quantum information
\begin{align}
Q^{(1)}(\mathcal{N})  = & \max_{\rho_S} I_c(\rho_S,\mathcal{N}), \\ \label{eqn:ic}
I_c(\rho_S,\mathcal{N}) =& S(\rho_{S^{'}}) - S(\rho_{RS'})= S(\rho_{RE^{'}}) - S(\rho_{E^{'}}),
\end{align}
where $S(\rho) = - \trace[\rho \log \rho]$ is the von Neumann entropy \cite{notelog}, $\rho_A$ is the reduced density matrix on $A$, and the maximum is taken over all density matrix ensembles $\rho_S$ on the system $S$.  An important identity that follows from Eq.~(\ref{eqn:ic}) is $S(\rho_S) - I_c(\rho_S,\mathcal{N}) = I(R:E{'}) \ge 0$, where $I(R:E{'})$ is the mutual information between the reference and the environment following the application of $U_{SE}$.  For input states such that $S(\rho_S) = I_c(\rho_S,\mathcal{N})$, the mutual information between the reference and the environment is exactly zero, which implies the existence of a perfect recovery operation according to the previous argument.   The single-use quantum channel capacity is, thus, the maximum possible amount of quantum information  that can be perfectly transmitted with repeated, independent uses of a noisy channel.  

In the theory of quantum error correction,  it is  helpful to generalize the single-use channel capacity to many copies using the limiting definition
\be
Q(\mathcal{N}) = \lim_{n \to \infty} \frac{1}{n} Q^{(1)}(\mathcal{N}^{\otimes n}) \ge Q^{(1)}(\mathcal{N}).
\ee
A finite value of $Q$ implies that approximate quantum error correction is possible on a space of asymptotic dimension $2^{n Q}$, even when it fails in the repeated use of a single copy of the channel.  However, this may come at the cost of performing  state preparation (encoding) and decoding operations on the $n$-fold replicated Hilbert space of the system \cite{Lloyd97,HolevoBook,Devetak05,Devetak05b}.  For the models considered below, one of our main results is the  explicit construction of optimal single-copy encoding operations, with an associated existence proof for single-copy decoding.

With the concept of channel capacity in hand, we define a purification phase transition with respect to a parameter $p$ (or set of parameters $\vec{p}$) in $\mathcal{N}_t$ and a partially protected subspace of $S$ of dimension $2^N$.  For $p<p_c$, and any  power-law relation between the effective number of qubits $N$ in the protected subspace and scaled time $t_{\bm{a}} = a_1 N^{a_2+1}+a_3(N)$ $(t_{\bm{a}} >0,a_2 > a_c)$ \cite{a3}, the channel capacity $Q_t$ of $\mathcal{N}_{t_{\bm{a}}}$ is extensive  $\lim_{N \to \infty} Q_{t_{\bm{a}}}/N  = c(p) > 0 $.   
  In contrast, for $p> p_c$, the channel capacity density converges to zero in this thermodynamic limit.    As a result, a purification phase transition can be interpreted as a type of error correction threshold for a family of protected subspaces acted on by a few-parameter family  $(\vec{p},t,N)$ of quantum channels $\mathcal{N}_t$.

For the random and uncorrelated channels considered below, the dynamics are statistically translationally invariant in time.  In this case, our definition of a purification transition above suggests a stronger condition that the subextensive corrections to the channel capacity are also time-independent on sufficiently long polynomial time scales $t>N^{a_c+1}$, i.e.,   for any sequence of allowed $\bm{a}(N)$ with $a_c < a_2$ there exists a sequence $\bm{b}(N)$ with $t_{\bm{a}} < t_{\bm{b}}$ such that $(t_{\bm{b}} - t_{\bm{a}})/N \to \infty$ and  $|Q_{t_{\bm{a}}} - Q_{t_{\bm{b}}}|$  converges to zero with $N$.  We refer to a purification transition satisfying this condition as a \textit{strong} purification transition.  

  For these definitions to apply to unitary-measurement models like the random Clifford model discussed above (where capacity achieving states are unknown), the definition of a purification transition has to be adapted  to the channel-averaged quantum capacity \cite{Choi19}
 \be \label{eqn:imax}
  \bar{I}_{\max} \equiv  \max_{\rho_0} \int d \sigma I_c(\mathcal{N}_\sigma, \rho_0),
  \ee
  where $d \sigma$ is the measure over the ensemble of random channels $\mathcal{N}_\sigma$.  The distinction from the quantum capacity is that the maximum over input states is taken after averaging the coherent quantum information over channels.  As pointed out in Ref.~\cite{Choi19}, for channel ensembles that are invariant under single-site unitaries, $\bar{I}_{\max}$ is maximized for the completely mixed input state taken in Sec.~\ref{sec:phases}. The basic argument is that the maximizing input state has to have  the same symmetries as the channel ensemble, which, in this case, implies that it is given by  the completely mixed state.   In Sec.~\ref{sec:traj}, we show that the coherent quantum information for unitary-measurement circuits is equal to the entropy of the mixed state averaged over measurement outcomes.  To strengthen the connections to our formal definitions, we present further numerical evidence in Sec.~\ref{sec:densitymat}  that the random Clifford model introduced in Sec.~\ref{sec:stab} realizes a strong purification transition with parameter $a_c \approx 1$ for any pair of sequences $(\bm{a},\bm{b})$ satisfying $a_c < a_2$ and $(t_{\bm{b}} - t_{\bm{a}})/ t_{\bm{a}} \to 0$. As a result, the channel-averaged maximum code rate $c(p) = S(\rho)/L$ can be approximated by the numerically observed values in Fig.~\ref{fig:model}(b) for $p < p_c$.

Our definition of a purification transition in terms of intrinsic properties of the quantum channel is somewhat distinct from the usual perspective on measurement-induced transitions, which are typically defined in terms of the properties of a particular ensemble of ``quantum trajectories'' \cite{Li18,Skinner18}.    To define a trajectory we express each $T_i$ in terms of a (not-necessarily unique) set of Kraus operators $ A_m $ 
  \be
  T_i(\rho) = \sum_{m} A_{m} \rho A_{m}^\dag,~\mathcal{N}_t(\rho) = \sum_{\vec{m}} K_{\vec{m}} \rho K_{\vec{m}}^\dag,
  \ee
  where $K_{\vec{m}} = A_{m_t} \cdots A_{m_1}$.  A quantum trajectory is given by an element of the ensemble $\{p_{\vec{m}},  K_{\vec{m}} \rho K_{\vec{m}}^\dag /p_{\vec{m}}\} $, where $p_{\vec{m}} = \trace[K_{\vec{m}}^\dag K_{\vec{m}} \rho]$ is the probability of a given measurement record $\vec{m}$.   Each trajectory has a physical interpretation in terms of a sequence of continuous time operations interspersed with generalized measurements acting on the system with measurement outcomes $\vec{m}$ \cite{CarmichaelBook,Plenio98}.  In Sec.~\ref{sec:traj}, we establish several concrete connections between the  trajectory viewpoint on measurement-induced transitions and the channel viewpoint taken in this work.  In particular, we show that when all possible choices of Kraus operators (i.e., unravelings)  lead to an ensemble of pure states  $K_{\vec{m}} \rho K_{\vec{m}}^\dag \propto \ket{\psi_{\vec{m}}} \bra{\psi_{\vec{m}}}$, then the system is in a ``pure phase'' ($p>p_c$).  
  On the other hand, when the system is in the ``mixed phase'' ($p<p_c$), then all possible unravelings of the dynamics result in the trajectories remaining mixed for exponentially long times.

The use of the replicated Hilbert space in the existence proofs for encoding and decoding pairs makes these operations difficult to study in specific models or implement in experiment.  To establish more explicit constructions of single-copy encoding and decoding operations, we introduce an additional constraint that also provides a natural connection between the channel and trajectory viewpoints:  We demand that each individual channel $T_i$ of $\mathcal{N}_t$ has at least one (not-necessarily unique) isometric embedding that evolves any two pure states of $S$ $\ket{\psi_{1,2}}$ as
\be \label{eqn:use}
U_{T_i} \ket{\psi_{k}}\ket{0} = \sum_{m } A_{m} \ket{\psi_{k }}  \ket{m }
\ee
where  $\bra{\psi_{k}} A_{m}^\dag A_{m'} \ket{\psi_{k'} } = \bra{\psi_{k}} A_{m}^\dag A_{m} \ket{\psi_{k'} } \delta_{m m'}$ and $\bra{m} m' \rangle = \delta_{m m'}$. This condition implies that there is a Kraus representation $\{ A_{m} \}$ for $T_i$, such that, for any input state $\rho$, $T_i(\rho) =  \sum_{m} A_{m} \rho A_{m}^\dag = \sum_{m} p_{m} \rho_{m}$ is given by a sum of orthogonal density matrices $\rho_{m} = A_{m} \rho A_{m}^\dag/p_{m}$ satisfying $\trace[\rho_{m} \rho_{m'}] \propto \delta_{m m'}$, where $p_{m} = \trace[A_{m}^\dag A_{m} \rho]$.  As a result, the channel $T_i$ always has at least one ensemble that can be unraveled by making projective measurements onto the image of the $A_m$. We call a channel satisfying Eq.~(\ref{eqn:use}) a \emph{unitary-dephasing} channel because it can be represented  by unitary dynamics followed by dephasing of some off-diagonal coherences in the density matrix.      

 For composite channels $\mathcal{N}_t$ formed from unitary-dephasing channels, there is a natural realization of quantum trajectories using what we call the \emph{monitored channels}
\begin{align}
\mathcal{N}_{\vec{m},t}(\rho) &= M_{m_t} \circ T_t \circ \ldots \circ M_1 \circ T_1(\rho), \\
M_{m_i}(\rho) & = P_{m_i} \rho P_{m_i}^\dag \otimes \ket{m_i}\bra{m_i},
\end{align}
where $P_{m_i}$ is an isometric projector onto the image of $A_{m_i}$ in a  representation of $T_i(\rho) = \sum_{m_i} A_{m_i} \rho A_{m_i}^\dag$ that satisfies Eq.~(\ref{eqn:use}).
The $\mathcal{N}_{\vec{m},t}$ are  completely-positive trace-nonincreasing maps that act on the input state as
\be
\mathcal{N}_{\vec{m},t}(\rho) = K_{\vec{m}} \rho K_{\vec{m}}^\dag \otimes \ket{\vec{m}}\bra{\vec{m}},
\ee
for $K_{\vec{m}} = A_{m_t} \cdots A_{m_1}$, thereby, directly realizing a quantum trajectory. We refer to the (trace preserving) complete mixture of monitored channels $\mathcal{N}_t^u = \sum_{\vec{m}} \mathcal{N}_{\vec{m},t}$ as an \emph{unraveled channel.} We define a set of monitored channels as having a (strong) purification transition if its unraveled channel has a (strong) purification transition in its quantum capacity or channel-averaged quantum capacity.     Note that unraveled channels have the special property that they are unitary-dephasing channels for every choice of $t$.

 Unitary-dephasing channels are examples of ``degradable'' quantum channels, which are defined by the condition that there is an isometric embedding for which $\rho_E$ can be obtained by a quantum channel acting on the system.   As shown by Devetak and Shor, the quantum channel capacity of a degradable channel is equal to the single-use channel capacity $Q(\mathcal{N}) = Q^{(1)}(\mathcal{N})$ \cite{Devetak05}.  This identity helps simplify the calculation of $Q$, but it does not provide explicit constructions for the encoding and decoding operations  operations or  imply that they can be done with repeated, independent use of a single-copy of the system \cite{gendeph}.  

Before continuing, we discuss an interesting connection between the unitary-dephasing condition in Eq.~(\ref{eqn:use}) and a foundational result in quantum error correction known as the Knill-Laflamme conditions \cite{NielsenChuang,Knill00}.  Given a set of pure quantum states, or ``codewords,'' $\ket{c_i}$ and a set of error operators $E_a$, the necessary and sufficient conditions for these errors to be correctable is 
\be \label{eqn:kl1}
\bra{c_i} E_a^\dag E_b \ket{c_j} = C_{ab} \delta_{ij}
\ee
where $C_{ab}$ is an arbitrary Hermitian matrix.   Changing into an orthonormal basis for $C$, $(V^\dag C V)_{ab} = \lambda_a \delta_{ab} $, implies that there is a special set of error operators $\bar{E}_a = \sum_{b} E_b V_{ba}$ that are uniquely identifiable
\be \label{eqn:kl2}
\bra{c_i} \bar{E}_a^\dag \bar{E}_b \ket{c_j} = \lambda_{a} \delta_{ab} \delta_{ij}.
\ee
If we interpret the Kraus operators in the quantum channel as possible error operators, then the unitary-dephasing condition in Eq.~(\ref{eqn:use}) essentially removes the constraint from the Knill-Laflamme conditions that Eqs.~(\ref{eqn:kl1})-(\ref{eqn:kl2}) are proportional to $\delta_{ij}$.  This weaker condition implies that the errors are detectable, even though the codewords are not protected.  In this setting, an intuitive picture for a strong purification transition in a unitary-dephasing channel is that, in the mixed phase, the system spontaneously projects the states of the system into codewords that restore the full Knill-Laflamme conditions \cite{sarang}.  As we show in Sec.~\ref{sec:codethm}, in general, this is achieved only approximately \cite{AKL,Beny10,Nielsen98,Barnum00}.   

Monitored channels naturally arise in the weak measurement picture for unitary-measurement dynamics that use ancilla qudits to perform the measurements of the system (see Sec.~\ref{sec:traj} and also Ref.~\cite{Choi19,Szyniszewski19,Jian19,Bao19}).  In this representation, the computational basis states of the ancilla are entangled with all possible paths in a quantum trajectory evolution and then ``dephased'' by an environment.  Thus, the quantum trajectory can be unraveled by an observer through projective measurements of the ancillae in their computational basis.   Several special properties of this class of channels were  exploited in recent works on the measurement-induced entanglement transitions \cite{Jian19,Bao19,Choi19}.  
 Here, we show that imposing the strong purification condition on a monitored channel is enough to prove that the monitored dynamics gives rise to an optimal encoding operations for the future evolution of the channel.

More specifically, we consider a monitored channel $\mathcal{N}_{\vec{m},t} = M_{m_t} \circ T_t \circ \cdots \circ M_{1} \circ  T_1$ that exhibits a strong purification transition.  We prove that $\mathcal{N}_{\vec{m},t}$ defines a family of  quantum error correcting codes that can be efficiently generated with a single use of $\mathcal{N}_{\vec{m},t_{\bm{a}}}$ for any allowed sequence of pairs $(\bm{a},\bm{b})$ in the thermodynamic limit.   These encodings have an associated  high-fidelity, single-copy recovery operation for the future evolution of the unraveled channel $\mathcal{N}^u_{t_{\bm{b}},t_{\bm{a}}} = \sum_{\vec{m}} M_{m_{t_{\bm{b}}}} \circ T_{t_{\bm{b}}} \circ \cdots \circ M_{m_{t_{\bm{a}}+1}} \circ T_{t_{\bm{a}}+1}$ on the polynomial time scale $t_{\bm{b}}- t_{\bm{a}}$ in the thermodynamic limit.  The amount of encoded information is optimal in the sense that it converges to the channel capacity of $\mathcal{N}_{t}^u$ for any $t_{\bm{a}} \le t \le t_{\bm{b}}$.  Our proof is based on randomly sampling codes generated by the channel dynamics that have the desired properties on average, which implies the existence of a large subset of the codes with the desired properties.

\section{Dynamically Generated Codes: Emergent Quantum Error Correction}
\label{sec:codethm}

In this section, we describe in greater detail the properties of the quantum error correcting codes in the mixed phase in both many-copy and single-copy quantum error correction protocols.   Our discussion of many-copy error correction follows directly from well-known existence proofs of encoding and decoding pairs that can achieve the channel capacity of any quantum channel in the asymptotic limit of an infinite number of copies \cite{Lloyd97,Devetak05b}.  The results on single-copy quantum error correction are specific to the class of unitary-dephasing channels introduced in the previous section and, to our knowledge, have not been previously discussed in the literature.

As we noted above, a key feature of a  purification transition is the associated existence of a quantum error correcting code on the replicated Hilbert space $\mathcal{H}^{\otimes n}$, which can be decoded on all polynomial timescales. To make this statement more precise, we let $\mathcal{N}_t = T_t \circ \cdots \circ T_1$ be a family of quantum channels indexed by integers $t$ with a purification transition.  We fix an $\epsilon >0$ and $\bm{b}$ with $1 < t_{\bm{b}}$.  Now, there is an $ N_{\epsilon}$ and $n_\epsilon$ such that, for all $N\ge N_\epsilon$ and $n\ge n_{\epsilon}$, $|Q_{t_{\bm{b}}}/N - c(p)| < \epsilon$ and there exists an encoding  $ \mathcal{E}_{\bm{b}}:\mathcal{H}_{n Q_{t_{\bm{b}}}} \to \mathcal{H}^{\otimes n}$ and a recovery (decoding) operation $ \mathcal{R}_{t_{\bm{b}}}: \mathcal{N}_{t_{\bm{b}}}^{\otimes n}(\mathcal{H}^{\otimes n}) \to \mathcal{H}_{n Q_{t_{\bm{b}}}}$ such that, for all input states $\rho$ \cite{Lloyd97,Devetak05b}
\be \label{eqn:rne}
|| \mathcal{R}_{0,t_{\bm{b}}} \circ \mathcal{N}_{t_{\bm{b}}}^{\otimes n} \circ \mathcal{E}_{\bm{b}}(\rho) - \rho || <  \epsilon,
\ee
where $||\rho -\sigma || = \frac{1}{2} \trace[|\rho -\sigma|] $ is the trace distance and $\mathcal{H}_n$ is a Hilbert space of dimension $2^n$.  The trace distance can be related to the mixed-state fidelity
\be \label{eqn:fid}
F(\rho,\sigma) = \max |\bra{\psi_\sigma} \psi_\rho \rangle |^2 \ge 1 - 2 ||\rho - \sigma ||,
\ee
where the maximum is taken over all purifications $\ket{\psi_{\rho}}$ and $\ket{\psi_{\sigma}}$ of $\rho$ and $\sigma$, respectively.  

It immediately follows from Eq.~(\ref{eqn:rne}) that this encoding operation can be successfully decoded  for all earlier times $t < t_{\bm{b}}$ using the recovery operation
\be
\mathcal{R}_{0,t} = \mathcal{R}_{0,t_{\bm{b}}} \circ \mathcal{N}_{t_{\bm{b}},t}^{\otimes n},
\ee
where $\mathcal{N}_{t_2,t_1} = T_{t_2} \circ \cdots T_{t_1+1}$.  In the case of a strong purification transition, these encoding and decoding pairs are optimal in the sense that they saturate the channel capacity  for any choice of $\bm{a}$ such that $t_{\bm{a}} \le  t_{\bm{b}} $ and $a_c < a_2,b_2$ in the thermodynamic limit.

The fidelity is a bound on the trace distance via Eq.~(\ref{eqn:fid}) and monotonically increases under quantum operations $F(\mathcal{N}(\rho),\mathcal{N}(\sigma)) \ge F(\rho,\sigma)$ \cite{Nielsen96}.  As a result, the encoded dynamics $\mathcal{N}_{t_2}^{\otimes n}\circ \mathcal{E}_{\bm{b}}$ at time $t_2 <t_{\bm{b}}$ can be approximately reversed to any point in time $t_1 \le t_2 $ with the operation
\be \label{eqn:r12}
\mathcal{R}_{t_1,t_2} =\mathcal{N}_{t_1}^{\otimes n} \circ \mathcal{E}_{\bm{b}} \circ \mathcal{R}_{0,t_{\bm{b}}} \circ \mathcal{N}_{t_{\bm{b}},t_2}^{\otimes n}.
\ee
Thus, within the encoded subspace, the dynamics of the channel acts as a type of effective, reversible unitary dynamics.

In the thermodynamic limit, we do not expect to require the infinitely replicated Hilbert space to construct high-fidelity encoding and decoding operations, provided one replaces the strong reversal condition in Eq.~(\ref{eqn:rne}) by a weaker, probabilistic reversal condition \cite{Barnum00,Barnum02}.  In this case, it is likely that any system exhibiting a strong purification transition can be encoded/decoded on the allowed polynomial timescales; however, we are not aware of a rigorous proof of this result. 
 Instead, we establish a version of this result for the special case of monitored channels $\mathcal{N}_{\vec{m},t}$ with a strong purification transition.  
 
 Monitored channels include all models known to exhibit measurement-induced entanglement transitions; however, whether such models generically exhibit a \emph{strong} purification transition has not been addressed. In Ref.~\cite{Choi19}, it was argued that a natural encoding operation for unitary-measurement dynamics is given by a high-depth random unitary circuit.  The general argument was based on the idea of quantum information scrambling in chaotic systems applied to quantum error correction problems \cite{Hayden07}, which, at a technical level, is related to the decoupling property of two-designs such as  (by definition) Haar random unitaries or the Clifford group \cite{Hayden07b,Brown12,Brown13}.  Although this encoding operation is provably optimal at vanishing measurement rates or finite times in the thermodynamic limit, the extension of this result to finite-measurement rates for times that scale polynomially in the system size was not addressed outside of the limiting case of nearly fully scrambling dynamics on large blocks of qubits.  
 
 More recent work has used replica methods to derive exact analytic mappings of the entanglement transition with Haar random two-site gates and single-site measurements to statistical mechanics models \cite{Bao19,Jian19}.  In the limit of infinite local Hilbert space dimension, these models map exactly to a certain percolation problem controlled by the measurement rate $p$ (see also Ref.~\cite{Skinner18} for a different mapping to percolation).  From this mapping, one can show that the system exhibits a  purification transition at the same value of $p$ as the entanglement transition in this limit \cite{Bao19}.  However, it is not currently known how to extend these analytic methods to the physically relevant case of finite local Hilbert space dimensions \cite{Jian19}, stabilizer circuit models like the random Clifford model, or more general random circuit models.  One promising direction for analytic progress is to consider dual-unitary random circuit models perturbed by measurements \cite{Kos20}. 
  
 Here, we show that imposing the strong purification transition condition is a strong enough constraint to imply that the encoding operation is directly implemented by the monitored channel dynamics.    Thus, in the mixed phase of a strong purification transition, the monitored channel dynamically generates a capacity-achieving quantum error correcting code for the future evolution of the channel.  We remark that this result was recently anticipated  in  Ref.~\cite{Fan20} (without making connections to channel capacities), where it was referred to as ``self-organized''  quantum error correction.   Similar to the existence of the single-copy encoding/decoding pair discussed above, this optimal code generation process may generically apply to systems exhibiting a strong purification transition; however, we are unaware of a rigorous proof for this conjecture.  Such a result would have broader implications for fault-tolerant quantum computing because, intuitively, it would suggest that the dynamics of a fault-tolerant system may generically project the system into an optimal code space for its future evolution.   We leave a more complete study of these possibilities for future work.
 We summarize our results on emergent quantum error correction in monitored channels in a  theorem:
  \begin{theorem} (Informal)
  \label{thm:code}
  Let $\mathcal{N}_{\vec{m},t} =M_{m_t} \circ T_t \circ \cdots \circ M_{m_1} \circ T_1$ be a monitored channel indexed by measurement outcomes $\vec{m}$ and integers $t>0$  with a strong purification transition.  There is an input state $\rho_0$ such that the density matrices $\rho_{\vec{m}} \propto \mathcal{N}_{{\vec{m}} ,t_{\bm{a}}}(\rho_0)$ obtained from evolving the monitored dynamics define optimal (capacity achieving) quantum error correcting codes for the future evolution of the unraveled channel $\mathcal{N}^u_{t,t_{\bm{a}}}=\sum_{\vec{m}} \mathcal{N}_{\vec{m},t,t_{\bm{a}}}$, i.e., $S(\rho_{\vec{m}} )/N \to c(p)$ and there is a high-fidelity reversal operation $R_{{\vec{m}}  t_{\bm{a}},t}$ such that the entanglement fidelity $F_e(R_{{\vec{m}}  t_{\bm{a}},t}\circ\mathcal{N}^u_{t,t_{\bm{a}}},\rho_{\vec{m}} ) \to 1$ in the thermodynamic limit for $t_{\bm{a}} \le t \le t_{\bm{b}}$.  Here, $c(p)$ is the channel capacity density of the unraveled channel $\mathcal{N}^u_t = \sum_{\vec{m}} \mathcal{N}_{\vec{m},t}$.
\end{theorem}
The formal statement and proof of the theorem are given in Appendix~\ref{app:proof1}.
This result is intuitively expected under the stated conditions, but the proof also provides useful bounds on the rate of convergence of the entanglement fidelity.  We show in Appendix~\ref{app:proof1} how to use these bounds, in combination with our numerical results in Sec.~\ref{sec:densitymat}, to determine when the random Clifford model can be approximated by a random unitary circuit for $p<p_c$.
  The methods of the proof are based on elementary arguments from approximate quantum error correction that ultimately rely on strong subadditivity of entropy \cite{Lieb73},  adapted to this class of monitored channels.
  The important role played by strong subadditivity can be understood from the general setup for the channel dynamics in Fig.~\ref{fig:channel}(c) in terms of three-components $R,$ $S$ and $E$.  The proof begins from a similar line of argument as we gave for many-copy quantum error correction.  In particular,  $\rho_0$ is chosen as a capacity achieving input state for the later time $t_{\bm b}$: $I_c(\rho_{0}, \mathcal{N}_{t_{\bm{b}}}^u) = Q_{t_{\bm{b}}}$.  
  
  In cases where such a state is difficult to find or initialize, one can generalize the notion of a strong purification transition to being dependent on a particular input state. The only requirement is that the coherent quantum information of this input state converges to an extensive value with time-independent subextensive corrections over some timescale.  The proof of the theorem above carries through to this case with only minor modifications.   In cases where the channel capacity density is zero, such as in the pure phase $(p>p_c)$, Theorem \ref{thm:code} still applies; however, when the channel capacity is not just subextensive, but strictly zero (or very close), then the proof of the theorem applies for the trivial reason that there is no code space where information can be stored and the recovery operation only needs to approximately succeed on a single-input state.  
    
  When the monitored channels $\mathcal{N}_{\vec{m},t}$ are drawn from a random ensemble (i.e., random channels), as discussed in Sec.~\ref{sec:pur}, we  define a channel-averaged strong purification transition with respect to a fixed input state $\rho_0$ by imposing the same convergence conditions on the channel-averaged quantum capacity $\bar{I}_{\rm max}$.   Since our proof is based on a Markov inequality, the theorem directly generalizes to the channel- and code-averaged code rate and recovery fidelity.  In models such as the random Clifford model, based on random unitary circuits interspersed with projective measurements \cite{Li18,Skinner18,Chan18b}, it is convenient to study channel-averaged strong purification transitions for the completely mixed input state $\rho_0 = \mathbb{I}/2^N$.   The completely mixed state is a natural input state for these models because its channel-averaged coherent quantum information often saturates $\bar{I}_{\rm max}$ (see Sec.~\ref{sec:pur}).

  We note that the entanglement fidelity $F_e(\mathcal{A},\rho) = F(\mathbb{I}_R \otimes \mathcal{A}(\ket{\Psi_{RS}}\bra{\Psi_{RS}}),\ket{\Psi_{RS}})$ is defined as the mixed state fidelity between a purification $\ket{\Psi_{SR}}$ of $\rho$ and its image under the channel $\mathcal{A}$   \cite{Schumacher96b}.  It is independent of the specific choice of $\ket{\Psi_{SR}}$ and quantifies the degree of quantum coherence of the channel.  For example, for any ensemble representation $e = \{\lambda_k,\ket{k} \}$ of $\rho = \sum_k \lambda_k \ket{k}\bra{k}$, $F_e$ is a lower bound on the average fidelity for state transmission $\bar{F}(\mathcal{A},e) = \sum_k \lambda_k F[\mathcal{A}(\ket{k}\bra{k}),\ket{k}] \ge F_e(\mathcal{A},\rho)$.  As a result, a direct consequence of our theorem is that randomly chosen input states from an arbitrary ensemble for $\rho_m$ can be evolved both forward and backward in time on all polynomial time scales in the thermodynamic limit \cite{note1}.   
  
  The recovery operations for intermediate times can be constructed in a similar manner to Eq.~(\ref{eqn:r12}) $(t_1 < t_2)$
  \be \label{eqn:rt1t2}
  \mathcal{R}_{m t_{1},t_2} = \mathcal{N}_{t_1,t_{\bm{a}}}^u \circ  \mathcal{R}_{m t_{\bm{a}},t_2}.
  \ee
This representation of the recovery operations relies on knowledge of the whole history of the evolution from time $t_{\bm{a}}$ to $t_2$ to recover back to time $t_1$. We can see more explicitly why this occurs by considering strong purification transitions where the subextensive corrections to $I_c(\rho_0, \mathcal{N}_{t_{\bm{a}}}^u) - I_c(\rho_0, \mathcal{N}_{t_{\bm{b}}}^u)$ decay to zero faster than $1/(t_{\bm{b}} - t_{\bm{a}})^4$ (see Appendix \ref{app:proof1}).  In this case, the decoding problem can be simplified by studying the action of the individual maps $T_i$ on the code space.   In particular, the action of the first channel on a purification of $\rho_m$ is 
  \be  \nonumber
  U_{T_{t_{\bm{a}}+1}} \ket{\Psi_{RSE}}  = \sum_{k \ell} \sqrt{p_{\ell |m}} \ket{k_{m}} A_{t_{\bm{a}}+1  \ell |m} \ket{\psi_{k | m}} \ket{\ell |m}.
  \ee 
According to Theorem \ref{thm:code}, each $A_{t_{\bm{a}}+1 \ell | m}$ can be approximated in the thermodynamic limit by an isometry into a set of orthogonal spaces indexed by $\ell | m$.   Let $\epsilon >0$ and $N$ be sufficiently large that the average recovery fidelity is $\bar{F}_e \ge 1 - \epsilon/(t_{\bm{b}} - t_{\bm{a}})$. The projective measurement operation $M_{\ell |m}$ lets us approximately map the dynamics to one of these isometries up to an average error $< \epsilon/(t_{\bm{b}}-t_{\bm{a}})$ \cite{noteisom}
\be \label{eqn:isometry}
  M_{\ell |m} \circ T_{t_{\bm{a}}+1} (\ket{\psi_{km}} \bra{\psi_{km}}) \to   U_{t_{\bm{a}}+1 \ell |m} \ket{\psi_{km}}, \\
\ee
for $\ket{\psi_{km}}$ sampled from any ensemble for $\rho_m$.   By an inductive argument, we can perform a similar mapping for each subsequent channel $T_t$ up until $t = t_{\bm{b}}$, while keeping the total average error bounded by $\epsilon$.

As a result, the evolution of the monitored channels $\mathcal{N}_{\vec{m},t}$, induces time-local unitary dynamics on the code space in the thermodynamic limit, with the usual group structure of unitary evolution.  However, the code space density matrix $\rho_m$ is continually evolving in a manner that depends on the measurement outcomes $\bm{\ell}_t = (\ell_{t_{\bm{a}}+1},\ldots,\ell_t)$.  Consequently, the unitary gates in each time step can depend on the full history of the previous gates applied to the system, which is consistent with the more general history dependence of Eq.~(\ref{eqn:rt1t2}).  We discuss further implications of these results in Sec.~\ref{sec:ftqc}.

\section{Monitored Random Circuits}
\label{sec:traj}

In this section, we discuss monitored channels defined in terms of random quantum circuits.  We also make more concrete connections between the monitored channel dynamics and the underlying dynamics of the composite unitary-dephasing channel.

We consider random quantum channels of the form
\begin{align} \label{eqn:Ntum}
\mathcal{N}_t(\rho) &=\sum_{\vec{m}} K_{\vec{m}} \rho K_{\vec{m}}^\dag \\
K_{\vec{m}} & = U_t P_t^{m_t} \cdots U_1 P_1^{m_1},
\end{align}
where $P_i^{m_i}$ is a sequences of positive-operator-valued-measures (POVMs) that  satisfy $\sum_{m} P_i^{m} = \mathbb{I}$,  $\vec{m}$ indexes the measurement outcomes, and $U_i$ are unitary operators that can depend on the most recent measurement outcomes.      
We will focus on the case where $m_i \in \{ 0,1\}$ takes one of two possible outcomes and $P_i^m$ are simple single-site projectors.  These random channels have a decomposition into unitary-dephasing channels as defined in Sec.~\ref{sec:pur}; however, whether the channel has a purification transition depends on the details of the unitary gates and the space-time position of the projectors.  In this work, we will consider the case where the projectors and unitaries all act on a fixed number of qubits.  Absent monitoring, such a channel will tend to drive the system towards infinite temperature. 

\begin{figure}[tb]
\begin{center}
\includegraphics[width = .4 \textwidth]{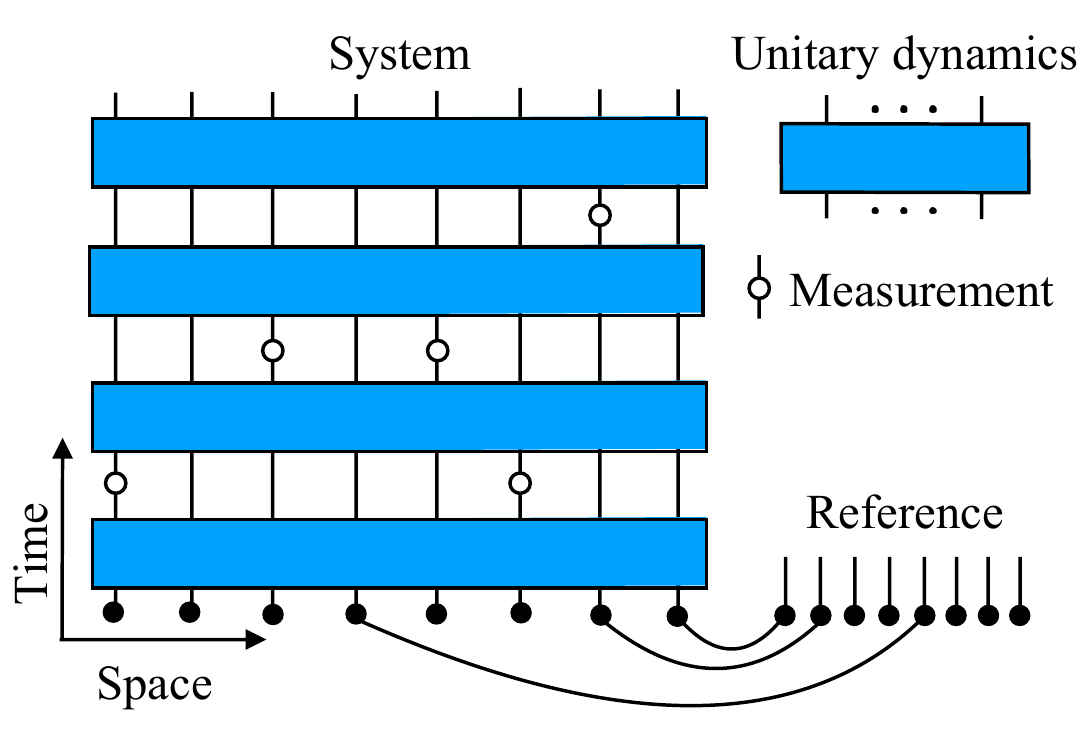}
\caption{Unitary-measurement dynamics in monitored random circuits.  The system is initially partially entangled with a set of reference qubits and undergoes unitary dynamics interspersed with measurements.  The two phases in a purification transition correspond to whether or not the measurements remove all coherent quantum information of the input state on polynomial time scales in the thermodynamic limit.   }
\label{fig:up}
\end{center}
\end{figure}

A natural unraveling into quantum trajectories $\{ p_{\vec{m}},\rho_{\vec{m}} \}$ for the channel dynamics takes the form 
\begin{align} \label{eqn:rhom}
\rho_{\vec{m}} &= K_{\vec{m}} \rho K_{\vec{m}}^\dag/p_{\vec{m}},~\mathcal{N}_t(\rho) =\sum_{\vec{m}} p_{\vec{m}} \rho_{\vec{m}},
\end{align}
where $p_{\vec{m}} = \trace[ K_{\vec{m}}^\dag K_{\vec{m}} \rho]$ is the probability of a given trajectory.   An equivalent description of the trajectory is in terms of a quantum circuit interspersed with measurements, as depicted in Fig.~\ref{fig:up}.

We show in Appendix~\ref{app:Ic} that the coherent quantum information of the full channel $\mathcal{N}_t$ is bounded by the average entropy of any trajectory ensemble 
\begin{align} \label{eqn:Ic}
I_c(\rho,\mathcal{N}_t) &\le \sum_{\vec{m}} p_{\vec{m}} S(\rho_{\vec{m}}).
\end{align}
This bound can be used to constrain whether the underlying channel $\mathcal{N}_t$ is in a mixed or pure phase.  To show this we need to consider the coherent quantum information of the replicated channel $\mathcal{N}_t^{\otimes n}$.
Applying the bound in Eq.~(\ref{eqn:Ic}) to $\mathcal{N}_t^{\otimes n}$ and using subadditivity of entropy results in the bound
\begin{align} \label{eqn:Icn}
I_c(\rho,\mathcal{N}_t^{\otimes n})& \le \sum_{i}  \sum_{\{\vec{m}_j\}} p_{\{\vec{m}_j \}}  S\big( \rho_{\{\vec{m}_j \} }^{(i)} \big), \\ \nonumber
\rho_{\{\vec{m}_j\}}^{(i)}& = p_{\{\vec{m}_j\}}^{-1} \trace_{\ell \ne i}[ K_{\{ \vec{m}_j \}} \rho K_{\{ \vec{m}_j \}}^\dag ], \\ \nonumber
K_{\{ \vec{m}_j \}} & = K_{\vec{m}_1} \otimes \cdots \otimes K_{\vec{m}_n},
\end{align}
where $\rho$ acts on the $n$-fold replicated Hilbert space and $\rho_{\{ \vec{m}_j\}}^{(i)}$ is the reduced density matrix of the $i$th replica conditioned on the measurement record across all $n$ replicas $\{\vec{m}_j \}$.   

Now, consider the case where all initial states converge to zero entropy density  averaged  along trajectories.  If we fix the measurement records on $\vec{m}_\ell$ on replicas $\ell \ne i$, then the density matrix $\rho_{\{\vec{m}_j\}}^{(i)}$ can be interpreted as a single-copy quantum trajectory that occurs with a conditional probability  $p(\vec{m}_i | \vec{m}_\ell,~\ell \ne i )$.  Averaging over these conditional probabilities implies that the average entropy density of $\sum_{\vec{m}_i} p(\vec{m}_i | \vec{m}_\ell,~\ell \ne i ) S(\rho_{\{\vec{m}_j\}}^{(i)})$ must also converge to zero for each value of the other measurement records.  As a result, each term in the upper bound in Eq.~(\ref{eqn:Icn}) converges to zero and the system is in a pure phase.  On the other hand, when some initial states have an extensive entropy averaged along trajectories, then the system can be in a mixed phase.    
   
These results show that the information theoretic properties of quantum trajectories strongly constrain the underlying quantum channel $\mathcal{N}_t$; however, it is important to note that channel and trajectory viewpoint are not equivalent.  In particular, there is a common intermediate case where some ensembles of quantum trajectories can remain mixed on average, while the underlying quantum channel is still in a pure phase (i.e., it has subextensive channel capacity) according to our definition.  To avoid these ambiguities and more formally investigate trajectory ensembles from the channel viewpoint, we consider the class of monitored and unraveled channels introduced in Sec.~\ref{sec:pur}
\begin{align} \label{eqn:Nt2}
\mathcal{N}_{t}^u(\rho) &= \sum_{\vec{m}}  \mathcal{N}_{\vec{m},t}(\rho), \\
\mathcal{N}_{\vec{m},t}(\rho)& = K_{\vec{m}} \rho K_{\vec{m}}^\dag \otimes \ket{\vec{m}} \bra{\vec{m}}
\end{align}
which stores the classical data associated with the measurement record in a register of ancillas.  Here, the input state $\rho$ is the reduced density matrix of a system of $N$ qubits without the ancilla and $\ket{\vec{m}}$ is the state of the ancilla qubits, which are assumed to always be initialized in the computational zero state $\ket{0}$.  
 The unraveled channel describes a physical evolution given by a sequence of unitary operations and perfect projective measurements of the system.  However, the ambiguity remains that even if one unraveled channel $\mathcal{N}_t^{m}$ is in a mixed phase, there is  often a different  unraveled channel $\mathcal{N}_t^{p}$ that will still be in a pure phase.  For example, such pairs of channels $\mathcal{N}_t^{m,p}$ arise in the monitored random circuits of the type studied in this work when one allows for arbitrary (potentially nonlocal and high weight) projective measurements of the system  in defining the unraveling \cite{Bao19}.   Given these subtleties, it is an interesting subject for future work to understand how, in the context of measurement-induced phase transitions,  different $\mathcal{N}_t^u$ are related to each other and to $\mathcal{N}_t$.  

It follows immediately from the definitions that the unraveled channel in Eq.~(\ref{eqn:Nt2}) is a unitary-dephasing channel of the type introduced in Sec.~\ref{sec:pur}. 
 For these unravelled channels,  the inequality in Eq.~(\ref{eqn:Ic}) is saturated \cite{choi}
 \be \label{eqn:Icequal}
I_c(\rho,\mathcal{N}_t^u) = \sum_{\vec{m}} p_{\vec{m}} S(\rho_{\vec{m}}),
\ee
We showed in Sec.~\ref{sec:stab} that the random Clifford model is a monitored channel with a purification transition.  We show below that it also satisfies the conditions needed for a strong purification transition.  Applying  Theorem \ref{thm:code} to this model then shows that, for $p<p_c$, the monitored channel dynamically generates a quantum error correcting code that protects against further loss of quantum information in single-copy quantum error correction.
 
  
More broadly, the dynamically generated codes for stabilizer circuits with a strong purification transition are examples of stabilizer quantum error correcting codes  \cite{Gottesman96}.  As a result, the associated encoding and recovery operations can be efficiently computed.   The code space varies according to the sequence of gates, measurement locations, and measurements outcomes in the circuit, but using knowledge of the gates and measurement locations, one can construct perfect encoding operations.  The recovery operations are  unitary Clifford circuits that require additional access to the measurement record to decode (e.g., see Appendix \ref{app:mixedevolve}).   
 There is also considerable numerical and analytical evidence that more general quantum trajectory models with nonstabilizer  dynamics undergo a strong purification transition \cite{Zabalo20,Bao19,Li20}.  For such models, it is not currently known whether the associated recovery operations can be efficiently implemented.

In spatially local models, it is natural to expect that the purification phase transition occurs  concurrently with the entanglement transition found recently in similar models \cite{Li18,Skinner18}. We find strong evidence for this scenario in the $1+1$ dimensional random Clifford model.  Several groups have also recently verified our results that these two critical points likely generally coincide in 1+1 dimensions without quenched disorder \cite{Bao19,Zabalo20,Li20}.  We discuss plausible conditions under which these critical points can move to different points in the phase diagram in Sec.~\ref{sec:phase}.

\section{Critical Properties of the Random Clifford Model in 1+1 Dimensions}
\label{sec:crit}

In this section, we perform a careful examination of the critical properties of the entanglement and purification phase transition of the random Clifford model in 1+1 dimensions.  We also discuss the properties of the optimal quantum error correcting code in the mixed phase and identify its strong purification phase transition parameters.  These parameters constitute an additional set of data that describe critical properties of the line of phase transitions in Fig.~\ref{fig:channel}(a) for $0\le p<p_c$.  This line of critical points is associated with the approach to the channel capacity limit in the mixed phase of the unitary-measurement dynamics.

\subsection{Entanglement Transition}

One of the central findings of our numerical study of this purification transition in $1+1$ dimensions is that it occurs concurrently, and with the same critical exponents, as the entanglement phase transition for pure initial states.  Thus, before examining the scaling behavior of the mixed state dynamics, we first revisit the critical properties of the entanglement phase transition for pure initial states.

\begin{figure}[tb]
\begin{center}
\includegraphics[width = .4 \textwidth]{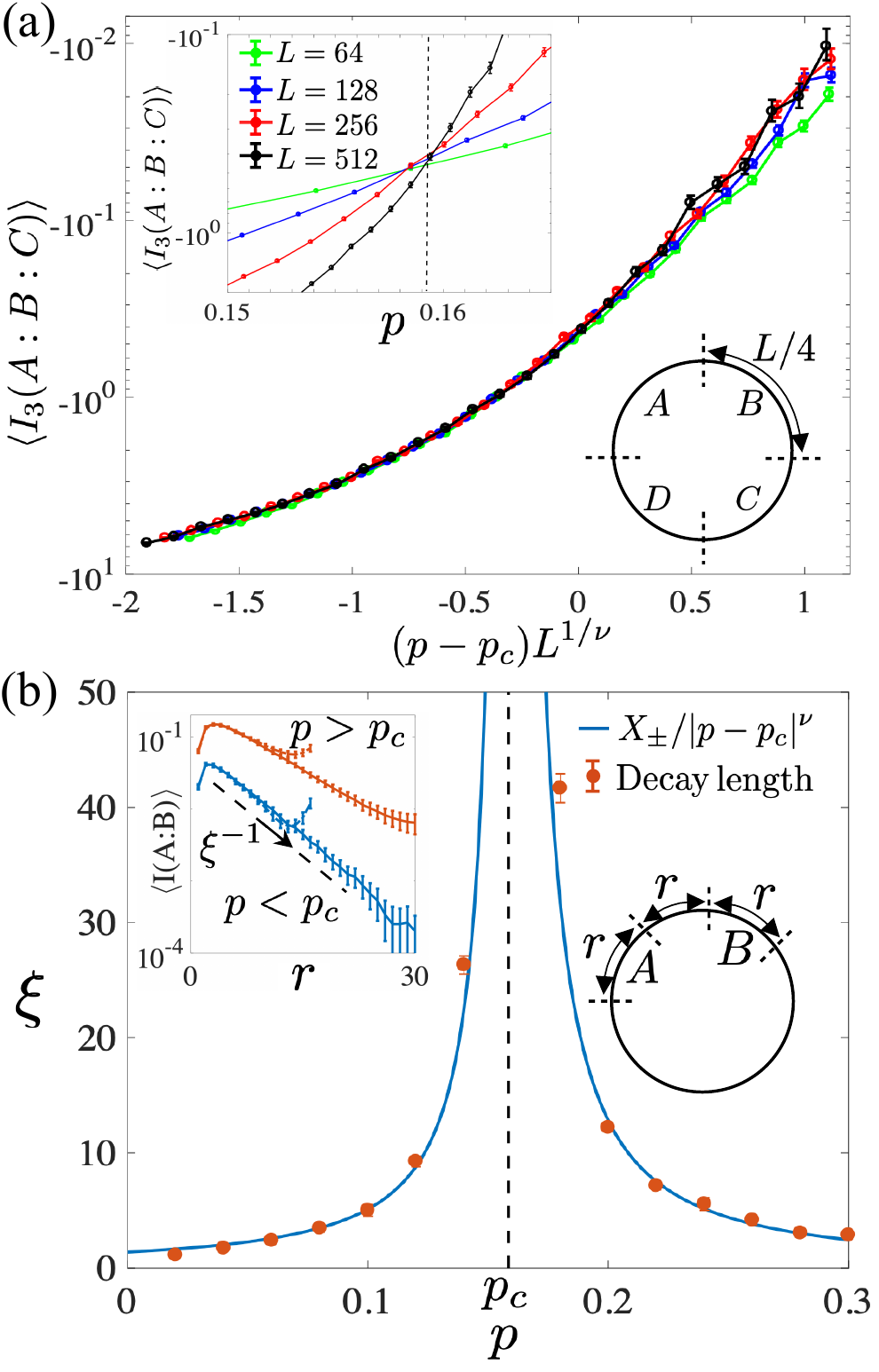}
\caption{(a) Average tripartite mutual information ${\mean{I_3(A:B:C)}}$ for three contiguous regions of length $L/4$ with pure initial states.  (inset) We can identify $p_c=0.1593(5)$ from the crossing point for different sizes. Through a collapse of the $L = 128-512$ data, we obtain $\nu = 1.28(2)$.  (b) Correlation length $\xi$ extracted from the decay with $r$ of $I(A:B)$ for $A$ and $B$ of equal size $r$ separated by a region of length $r$.  $\xi$ is well fit by the function $X_\pm/|p-p_c|^\nu$ with $X_\pm = 0.18/0.12$ for $p \gtrless p_c$. (inset) Sample of the data used to extract $\xi$ for $p\gtrless p_c$, $L = 64$ and $128$, and  $t = 4L$.  Completely-mixed initial conditions were chosen to avoid a a strong feature that appears near $r = L/4$ for pure initial states with $p<p_c$. }
\label{fig:MI}
\end{center}
\end{figure}

 In Ref.~\cite{Li19}, it was shown that the critical region of the entanglement phase transition can be precisely identified by looking at the  mutual information  $I(A:B) = S(\rho_A) +S(\rho_B) - S(\rho_{A\cup B})$ between two antipodal regions on the circle of length $L/8$.   We have found that a similar, but more accurate, probe of the critical point is to use the tripartite mutual information between 3 contiguous regions of length $L/4$  [see inset to Fig.~\ref{fig:MI}(a)], defined as $I_3(A:B:C) = I(A:B)+I(A:C)-I(A:BC)$.  For pure states, $I_3(A:B:C)$ is symmetric under all permutations of $(A,B,C,D)$, where $D$ is the rest of the sample.  The tripartite mutual information, sometimes referred to as the topological entanglement entropy \cite{Kitaev06,Levin06}, is a natural measure of the degree to which information in a quantum wavefunction is encoded nonlocally.  Similar to the $L/8$ antipodal mutual information, it has the effect of removing the logarithmic divergences that appear in the halfcut entanglement at the critical point, which reduces finite-size corrections to scaling.  We find that $\mean{I_3(A:B:C)}$ vanishes in the pure phase, approaches a universal constant $\sim -0.5$ at the critical point, and has the expected negative volume-law scaling in the mixed phase.  This behavior can be explained using a minimal cut picture for the entanglement scaling near the critical point \cite{Zabalo20}.  In Fig.~\ref{fig:MI}(a), we show the finite-size scaling near the critical point for $I_3$ starting from pure initial states.  We observe a clear crossing of $I_3$ vs. $p$ with increasing $L$, which we use to identify $p_c = 0.1593(5)$.  From collapsing the $L = 128 - 512$ data with this value of $p_c$, we extract  $\nu = 1.28(2)$.  These estimates are consistent with those reported in Ref.~\cite{Li19}.
 
Another basic quantity of interest in characterizing the transition is the correlation length $\xi \propto |p-p_c|^{-\nu}$.   To realize a quantitative measure of the correlation length on both sides of the transition, we study $I(A:B)$ for the region shown in an inset of Fig.~\ref{fig:MI}(b), where $A$ and $B$ are of equal size $r<L/4$ and separated by a region of length $r$.  The mutual information is a basis independent correlation metric that can be used to upper bound all connected correlation functions between $A$ and $B$ \cite{Wolf08}.   This quantity decays exponentially with $r$  with a decay length that diverges as $p \to p_c$.  To reliably extract the decay length we found that it is better to begin with the completely-mixed initial state and run only until time $t=4L$ because pure states for $p<p_c$ develop a strong feature near $r = L/4$, which obscures the exponential decay.  This feature for pure initial states arises because of the volume-law entanglement, which implies that, when $r$ exceeds $L/4$, ${ I(A:B)} $ grows linearly with $r$ due to the volume-law scaling that appears for $L/4 < r \leq L/3$. 
Essentially, $r\geq L/4$ is the regime where $(A \cup B)^c$ is not larger than $A \cup B$, so it is an inadequate bath to fully decorrelate $A$ and $B$.  In contrast, for mixed initial states, the bipartite mutual information grows sublinearly in time, leading to a much weaker feature near $r = L/4$ for early times.

\subsection{Purification Dynamics}

\begin{figure}[tb]
\begin{center}
\includegraphics[width = .4 \textwidth]{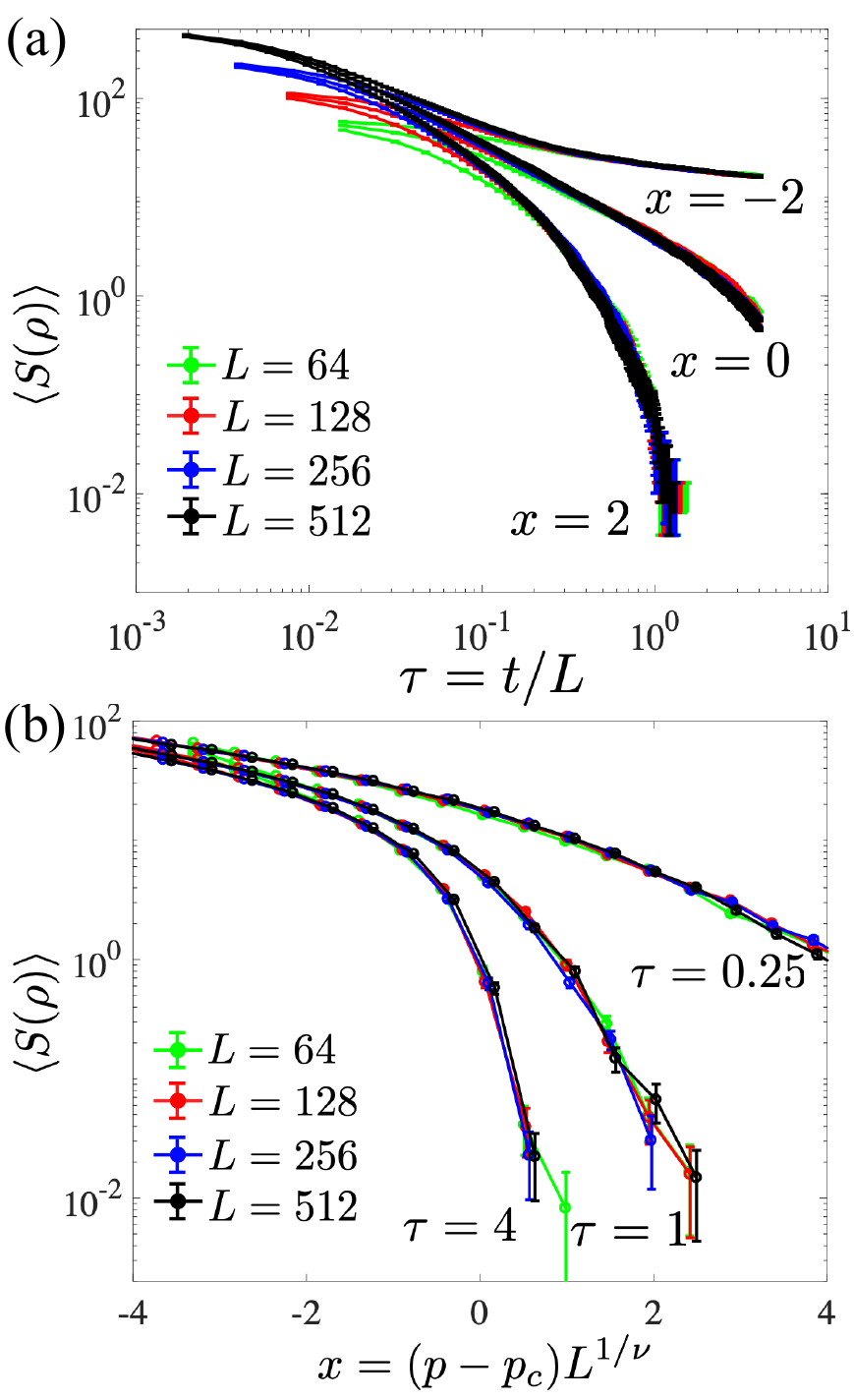}
\caption{(a) Average entropy $\mean{S(\rho)}$ vs.\ scaled time $\tau$ in the mixed phase $p< p_c$, at the critical point $p= p_c$, and in the pure phase $p>p_c$. (b) Finite-size scaling of $\mean{S(\rho)}$ across the transition for different values of the scaled time.}
\label{fig:purification}
\end{center}
\end{figure}

With the basic, static scaling properties of the entanglement transition established, we can now turn to the dynamical scaling of the purification transition.  The central results are summarized in Fig.~\ref{fig:purification},  which shows the scaling behavior of the average entropy across the transition.  Because of the apparent conformal symmetry at the critical point of this model \cite{Li19}, we assume a dynamical critical exponent $z= 1$ and take a dimensionless scaling function for the entropy
\be \label{eqn:srho}
\mean{S(\rho)} = F( x, \tau),
\ee
where $x = (p-p_c)L^{1/\nu}$ and $\tau = t /L$.  Note, that the scaling dimension of the entropy is zero, such  that  the $L$-dependence of $\mean{S(\rho)}$  enters only through the scaling function.  This is consistent with the behavior we observed recently for a different class of open-system entanglement phase transitions \cite{Gullans19b}.  This scaling has also recently been verified using arguments based on conformal symmetry \cite{Li20}.  

 In Fig.~\ref{fig:purification}(a), we show the scaled time-dependence of the entropy across the phase transition.  For $p > p_c$, there is rapid exponential decay of the entropy.  At the critical point $p=p_c$, there is an intermediate time regime during which we observe power law decay of the form $F(0,\tau) \sim 1/\tau$, which then crosses over to an exponential decay as the entropy approaches 1 bit.  For $p < p_c$, we see an initially rapid decrease in the entropy, which crosses over towards an exponentially slow decay at late times.  In Fig.~\ref{fig:purification}(b), we plot the entropy across the phase transition for different values of the scaled time.  In all cases, we see an excellent collapse of the data for $L$ ranging from 64 to 512.

\subsection{Structure of Optimal Code Space Density Matrices}
\label{sec:densitymat}

As shown in Theorem \ref{thm:code} in Sec.~\ref{sec:codethm}, the late-time density matrix  defines an optimal quantum error correcting code space for the channel dynamics, which motivates the study of the structure of its correlations.  In addition, these studies help make a more direct comparison between the purification phase transition for mixed initial states and the entanglement phase transition seen for pure initial states.

\begin{figure}[tb]
\begin{center}
\includegraphics[width = .4 \textwidth]{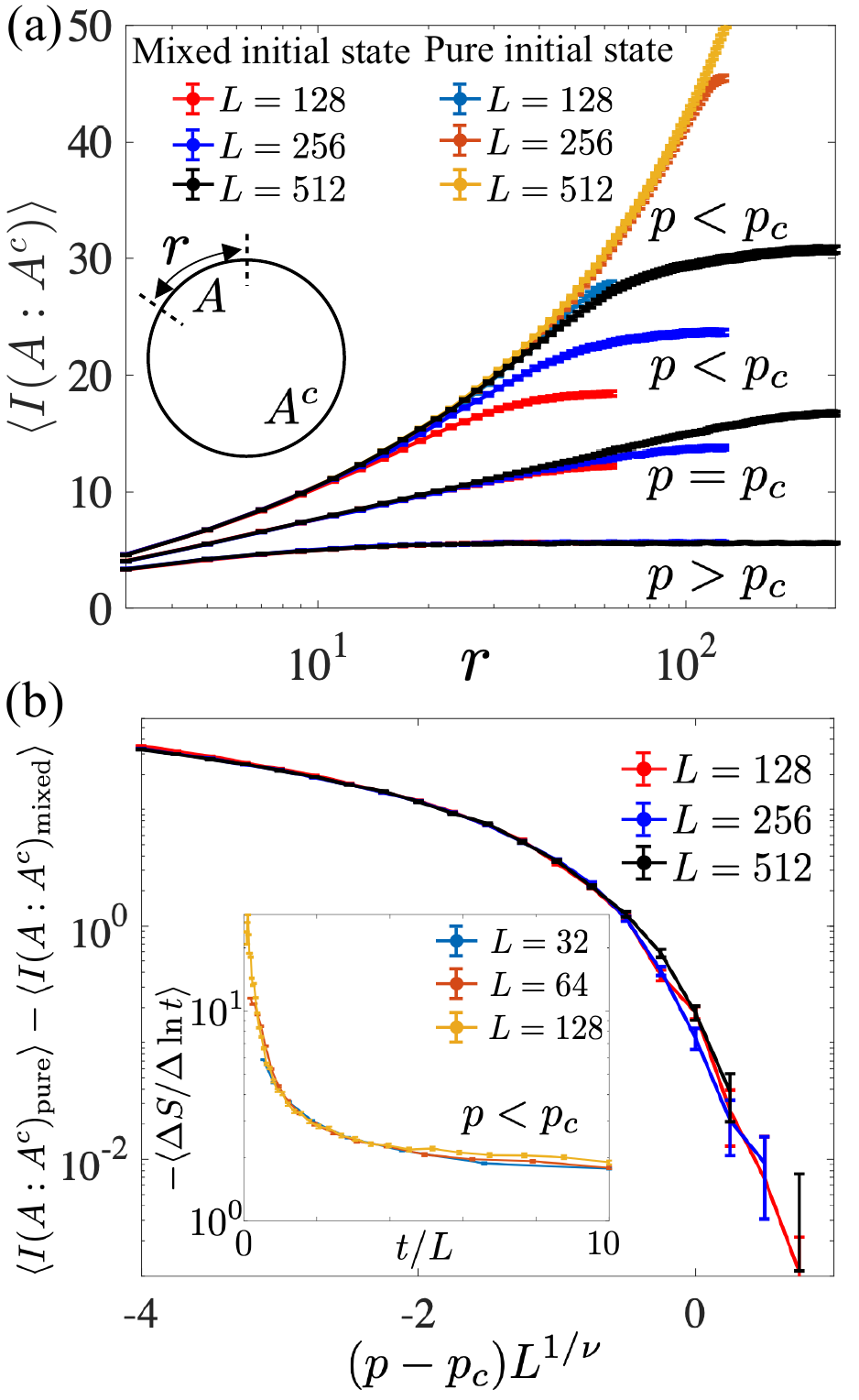}
\caption{(a) Finite-size scaling of the average bipartite mutual information $\mean{I(A:A^c)}$ at $t=4L$ with $A$ a contiguous region of length $r$, and $A^c$ its complement.  In the pure phase ($p=0.20>p_c$ is shown) and at the critical point, $\mean{I(A:A^c)}$ becomes independent of initial conditions, displaying area-law behavior in the pure phase and logarithmic scaling with $r$ at the critical point.  In the mixed phase ($p=0.12<p_c$ is shown), $\mean{I(A:A^c)}$ strongly diverges between mixed and pure initial conditions, displaying volume-law scaling for pure initial states and sublinear scaling for mixed states on this time scale.    (b) Finite size-scaling of the difference of halfcut mutual information between pure and completely-mixed initial conditions at $t=4L$.   (inset) Logarithmic time derivative of the entropy in the mixed phase $(p = 0.08 <p_c$ is shown), which has a power-law decay $\sim 1/t$ for $t \ll L$ and approaches a constant $L$-independent value at late times $t \gg L$.  The constant logarithmic derivative at late times is consistent with an exponentially long purification time. }
\label{fig:ent}
\end{center}
\end{figure}

 A convenient diagnostic of the mixed state density matrix is the bipartite mutual information $I(A:A^c)$ for $A$ of varying length $r$.  For pure states, $I(A:A^c)$ reduces to twice the entanglement entropy.  The main qualitative features are illustrated in Fig.~\ref{fig:ent}(a).  For $p \ge p_c$, the mixed state purifies on a timescale at most linear in $L$, which implies that pure and mixed initial conditions have identical late-time scaling behavior for $I(A:A^c)$.  In the pure or area-law phase, the long-time states exhibit only area-law mutual information, which quickly converges to a constant value independent of $L$.  At the critical point, the system builds up logarithmic mutual information.  The most significant result presented here is that, for $p< p_c$, we observe a sublinear scaling of $I(A:A^c)$ for completely-mixed initial states in contrast to the volume-law scaling observed previously for pure-initial states.  The presence of a logarithmic background for stabilizer circuits was identified through an indirect measure in Refs.~\cite{Li19,Chan18b}.  We find that mixed state dynamics partially reveals this feature by stripping away the volume-law mutual information. 

 Since the mixed state recovers the subextensive contributions to the bipartite mutual information for $p\le p_c$, a natural approach to finite-size scaling of the transition is to look at the difference of $\mean{I(A:A^c)}$ between pure and completely-mixed initial conditions.  The results are shown in Fig.~\ref{fig:ent}(b), where we see an excellent scaling collapse of the data using the value of $p_c$ and $\nu$ obtained from Fig.~\ref{fig:MI}(a).   At the critical point,  where the system purifies after a time $\sim L$ and builds up logarithmic mutual information $\mean{I(A:A^c)} \sim 2 \alpha(p_c) \ln L$, we find $\alpha(p_c) \approx 1.63(3)$ \cite{Li19}. Away from the critical point, it is difficult to reliably extract $\alpha(p)$ from the mutual information of the mixed state.

 According to our definition of a strong purification transition, the late-time behavior of the subextensive corrections to the coherent quantum information  are crucial in determining the efficiency of the encoding operation.  We present a finite-size scaling analysis of this dynamics for the completely mixed initial state in the inset to Fig.~\ref{fig:ent}(b) for $p = 0.08 < p_c$.  We find a consistent scaling collapse using the ansatz
 \be
 - \mean{\Delta S/ \Delta \ln t} = F(t/L),
 \ee
  with the asymptotic behavior $F(t/L) \sim L/t$ for $t \ll L$ and $F(t/L) \sim {\rm const.}$ for $t\gg L$.  Similar to Eq.~(\ref{eqn:srho}), we find that the scaling dimension of the entropy is zero and the $L$-dependence enters only through the scaling function. The late time behavior is consistent with an exponentially long purification time.  The early-time behavior exhibits a power-law scaling of the same form as the early-time behavior of $\mean{S(\rho)}$ at the critical measurement rate $p_c$.  Connecting to our definition of a strong purification transition, this power-law decay further implies that $a_c \approx 1$ for this strong purification transition, while the constant logarithmic derivative at late times requires the pair of sequences $(\bm{a},\bm{b})$ to satisfy $(t_{\bm{b}} - t_{\bm{a}})/t_{\bm{a}} \to 0$ in the thermodynamic limit.  These two conditions appear specific to the optimal code generation process because starting with an entropy density significantly below the channel capacity limit leads to a much more rapid convergence (not shown) to the plateau value.  For these input states that do not maximize the channel-averaged quantum capacity, the likely strong purification parameter is $a_c \approx 0$ for any pair of sequences $(\bm{a},\bm{b})$ with $0<a_2$ and $t_{\bm{a}} < t_{\bm{b}}$ in the thermodynamic limit.

We remark that recent work by Fan, Vijay, Vishwanath, and You obtained some characterizations of the dynamically generated  quantum error correcting codes identified in this work   \cite{Fan20}.  They focused on the mixed/volume-law phase of a more general Haar random model (i.e., the two-site Clifford gates are replaced by Haar random gates) in one spatial dimension, but many of their arguments can also apply to the Clifford model and to other geometries.  Using an approximate mean-field theory (see also Ref.~\cite{Shtanko20}), they found that the late-time entanglement of a pure state for $p<p_c$ always has a background logarithmic contribution $\sim \alpha_{\rm MF} \ln L$ ($\alpha_{\rm MF} = 3/2$), qualitatively consistent with the numerically observed behavior for stabilizer circuits shown in Fig.~\ref{fig:ent}(a).   This background logarithm term was argued to be a crucial aspect of the inherent encoding of the initial state of the system into a quantum error correcting subspace.  Our results provide additional evidence in support of this scenario.  Theorem \ref{thm:code} combined with arguments presented in Ref.~\cite{Choi19} imply that starting from the completely mixed state generates the optimal stabilizer code for this dynamics.  We have shown that this optimal code space density matrix is characterized by sublinear average mutual information.     Thus,  the background subextensive corrections to the pure state entanglement are likely an intrinsic aspect of the optimal quantum error correcting code generated by the dynamics.  
 
 In addition, Fan \textit{et al.} argued that one can bound the critical measurement rate ($p_c \le 0.1893$ for qubits) by arguing that the states in the volume-law phase are effectively encodings of a random Page state in a nondegenerate quantum error correcting code.  Although they present the argument for one-dimensional models, the nondegenerate code bound would apply to other geometries with two-site gates followed by measurements of each qubit with probability $p$.  Note, a degenerate code is one that can correct more errors than it can uniquely identify (e.g., via syndrome measurements for stabilizer codes) \cite{NielsenChuang}.   In stabilizer codes, a sufficient condition for a code to be degenerate is that its stabilizer group has elements of weight less than the code distance (minimal weight of the logical operators not in the stabilizer group).  Short elements in the stabilizer code group are intrinsic to unitary-measurement dynamics because the measurements are continually ``injecting'' single-site operators into the code-space density matrix. Therefore, the significance of this bound on $p_c$ for this problem is not immediately clear.  We  present evidence below that it is a relevant bound in $1+1$ dimensions, as argued by Fan \textit{et al.}, but not necessarily for higher dimensions as we observe an explicit violation in an all-to-all model.
  
  A key property of a degenerate code is that it can still have an error correction threshold even when the distance  does not scale extensively in the system size.  We can sensitively test this aspect of the codes in the volume-law phase because we can generate the optimal ensemble of codes with respect to the code rate.   By computing a quantity we call the contiguous code length, we show in Appendix \ref{app:codelength} that the average distance of the optimal  codes in 1+1 dimensions is subextensive for $p<p_c$, but apparently $\sim L$ as one approaches the critical point.  As a result, the bound put forth in Ref.~\cite{Fan20} may accurately apply to 1+1 dimensional systems near $p_c$, even though the optimal codes deep in the volume-law phase appear to be highly degenerate.  
 
 Outside 1+1 dimensions, it is natural to expect that the code distance no longer has to scale linearly with the number of qubits near $p_c$.  In the next section, we  study an all-to-all generalization of this 1+1 dimensional stabilizer circuit model,  finding $0.30 < p_{cp} \le 2/3$ for the purification critical point $p_{cp}$, in strong violation of the nondegenerate code bound.   As a result, even the codes near the critical point in more than one dimension can be highly degenerate.     As we discuss in Sec.~\ref{sec:disc}, these channel-averaged optimal codes are of potential practical relevance to fault-tolerant quantum computation, which motivates a more detailed understanding of their properties and performance.

\section{Mixed Phase in All-to-All Models}
\label{sec:bob}

 In this section, we present a basic demonstration of the existence of a mixed phase  in all-to-all models.  We  show that the purification critical point $p_{cp}$ violates a version of the Hamming bound for nondegenerate codes that may apply in 1+1 dimensions \cite{Fan20}.  
 
\begin{figure}[tb]
\begin{center}
\includegraphics[width = .42 \textwidth]{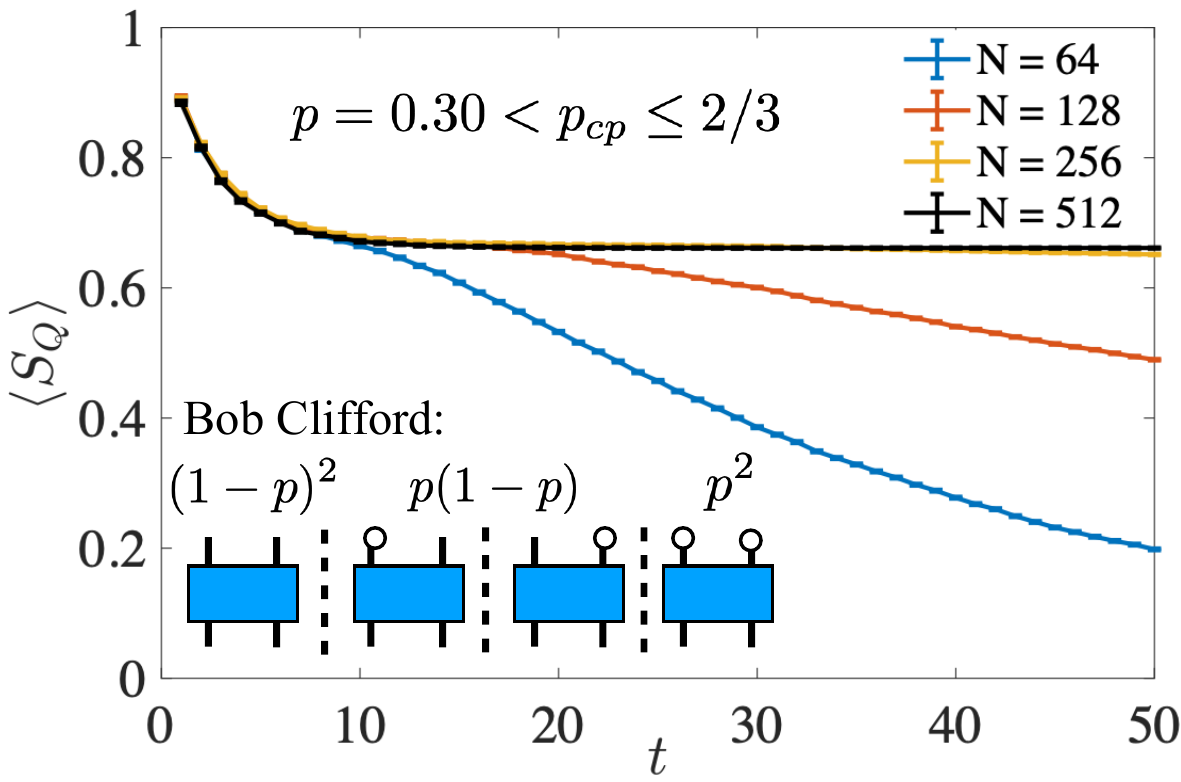}
\caption{ Entropy of the system entangled with a single reference qubit vs. circuit depth starting from the reference qubit in a maximally entangled state with one site and the rest of the system in a random stabilizer state.  At each time step $\Delta t = 2/N$ a pair of qubits is randomly selected and one of the four operations in the inset is applied.    The late-time entropy density serves as an order parameter for the mixed phase \cite{Gullans19e}.  It approaches a plateau value that it stays near for a time that increases exponentially in $N$ over this range of sizes. This provides an estimated lower bound on $p_{cp}$, while an upper bound of $p_{cp} \le 2/3$ can be obtained from a percolation mapping for this circuit \cite{Skinner18,Gullans19c,Nahum20}.  }
\label{fig:bobcliffsq}
\end{center}
\end{figure}

 We consider random circuit models as above with two-qubit gates and single-site measurements, but the gates are allowed to act on arbitrary pairs of qubits in the system that are chosen at random.   In the context of measurement-induced transitions, the existence of a phase transition has not yet been established in all-to-all models.   We also call these all-to-all models ``Bob,'' which is an acronym for ``bag-of-bits'':  Each time you apply a gate, you reach in to the bag of qubits and pull out two of them at random to act on with a gate (see  Fig.~\ref{fig:bobcliffsq}).

 In these Bob models, the basic notion of an area-to-volume-law entanglement transition (at $p_{ce}$) has to be revisited because there is no clear distinction between surface and bulk in this geometry: All qubits are adjacent to any entanglement cut.  On the other hand, there is still a natural anisotropy between space and time in such models.  Since a purification transition is fundamentally about memory of initial conditions, the definition in Sec.~\ref{sec:pur}   naturally generalizes to this case.   Entanglement transitions refer to the geometric structure of the correlations in the system and require more care to properly define in this setting.  As a result, we defer a more detailed study of measurement-induced phase transitions in all-to-all models to future work \cite{Gullans19c}.
 
To establish the existence of the mixed phase, we turn to the Bob Clifford model illustrated in the inset of Fig.~\ref{fig:bobcliffsq}.  At  each time step,  we randomly select a pair of sites, apply a random two-qubit Clifford gate, and measure each of the two sites with probability $p$.  In the case of Haar random gates, one can show analytically using the mapping to a percolation problem provided by Skinner, Ruhman, and Nahum \cite{Skinner18} that this model undergoes a percolation or connectivity transition in the Hartley entanglement entropy at $p_{cc} = 2/3$ \cite{Gullans19c,Nahum20}.  For $p>2/3$, the late time density matrix is necessarily in a zero entropy pure state regardless of the choice of initial conditions or unitary gates, thus, $p_{cp} \le 2/3$ for this model.

We obtain a lower bound on $p_{cp}$ by using the order parameter for the mixed phase introduced in our recent work \cite{Gullans19e}.  The distinction from the discussion above is that we replace the reference system that scales extensively with the system size by a single reference qubit.  For each circuit, we define
\be
S_Q = \sum_{\vec{m}} p_{\vec{m}} S(\rho_R),
\ee
as the entropy of this reference qubit averaged over trajectories.    In the mixed phase, the channel-averaged $S_Q$ will persist for exponentially long times.  

The qualitative behavior of $\mean{S_Q}$ in the mixed phase is shown in Fig.~\ref{fig:bobcliffsq} for $p = 0.30$.  To set up these simulations, we choose an initial state given by a random Clifford state, we then measure one site in the system, and place this qubit in a maximally entangled state with the reference qubit.  Similar to the purification dynamics of the completely mixed state, the average entropy of the reference qubit purifies for a short period that is independent of $N$ before crossing over to a late time plateau.  The persistence time of this plateau  diverges exponentially in $N$ over this range of sizes. The results of these numerics indicate that $0.30 < p_{cp} \le 2/3$ strongly violates the nondegenerate code bound for 1+1 dimensional qubit models $p_{cp} \le 0.1893$ put forth in Ref.~\cite{Fan20} (see discussion in Sec.~\ref{sec:densitymat}).  The optimal codes near the critical point are likely degenerate, making them potentially useful for applications in fault-tolerance due to the higher error thresholds that are possible with zero-rate, degenerate quantum codes, such as the surface code  \cite{Dennis02}.

 \section{Discussion}
 \label{sec:disc}

 \subsection{ Phase Diagram and Universality Classes} 
\label{sec:phase}

\begin{figure}[tb]
\begin{center}
\includegraphics[width = .49 \textwidth]{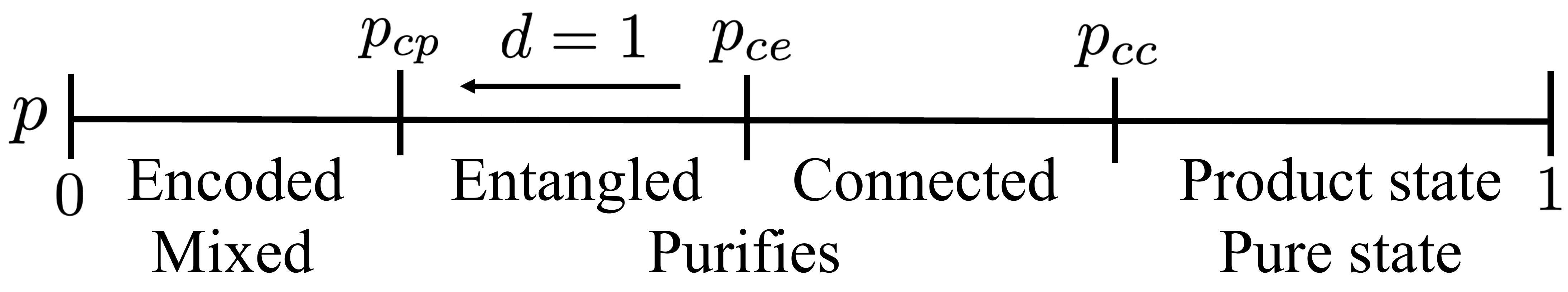}
\caption{General phase diagram for unitary-measurement models.  At large values of $p$, the system breaks up into product states over disconnected finite clusters.  As $p$ is lowered, the system first undergoes a connectivity or percolation transition at $p = p_{cc}$ where an infinite cluster in the quantum wavefunction forms \cite{Skinner18}.  Once the infinite cluster forms it can undergo a phase transition from subextensive to extensive bipartite entanglement at $p = p_{ce}$ \cite{Li18,Skinner18,Li19}.  At the purification transition point $p_{cp}$, we showed here that quantum information in the system becomes reliably encoded in a finite-rate, capacity achieving quantum error correcting code. In 1+1 dimensions without quenched disorder, the purification and entanglement transitions appear to generally occur at the same point in the phase diagram.  The appropriate classification of these phase transitions with quenched disorder or in higher dimensions  remains open.}
\label{fig:phasediaggen}
\end{center}
\end{figure}

A phase diagram for the broader family of entanglement and purification phases in unitary-measurement models  is presented in Fig.~\ref{fig:phasediaggen}.  Part of the richness of these models arises from the possibilities for multiple phase transitions with  intermediate phases.  For example, in $1+1$ dimensional Haar models, the first entanglement transition that occurs as one lowers the measurement rate from $p=1$ is a connectivity or percolation transition where the wavefunction sharply changes from a perfect product state over finite clusters to a state where one of those clusters is extensive (we call this transition point $p_{cc}$) \cite{Skinner18}.  A useful diagnostic for $p_{cc}$ in this case is the Hartley entanglement entropy, which has an exact analytical mapping to a percolation problem. This connectivity transition has also been found to occur in close proximity to a transition from Poisson to Wigner-Dyson level-statistics in the entanglement spectrum for pure states  \cite{Zhang20}.  For stabilizer circuit models, the Hartley entropy is not a good diagnostic for $p_{cc}$ because the entanglement spectrum is always degenerate.

For Haar-random circuits with brickwork geometry in 1+1 dimensions, the entanglement transition for the von Neumann entropy occurs at a much lower value ($p_{ce} \approx 0.17$) than the analytical result for 2D percolation $p_{cc} = 1/2$  \cite{Skinner18,Zabalo20}.  Interestingly, Bao, Choi, and Altman found a whole family of critical points at intermediate values of $p$ between $p_{ce}$ and $p_{cc}$ \cite{Bao19}; however, these critical points emerged only after weighting the trajectories by powers of their Born probability and circuit probability, which leads to the averages being dominated by rare realizations that may not reflect the behavior of  typical realizations in the thermodynamic limit.

In this work, we have provided a stringent definition of purification transitions in terms of the formation of an extensive quantum error correcting code space in the system.  Calling this purification transition point in the phase diagram $p_{cp}$, the behavior for $p<p_{cp}$ has many of the properties  associated with the $p = 0 $ limit of random unitary dynamics.  In particular, we showed in this work that the late-time dynamics acts as a type of random unitary circuit on an extensive encoded subspace, which will generically generate extensive entanglement.  As a result, we suspect that the purification critical point is the most stringent type of ``ordering'' in these models.  In the 1+1-dimensional model studied here, we found that $p_{ce} = p_{cp}$, which has now been argued to generally hold in 1+1 dimensions without quenched disorder \cite{Bao19,Li20}. 

In future work, it will be interesting to determine if there are any conditions under which $p_{cp}$ and $p_{ce}$ can differ from each other.  In quantum trajectories, entanglement ordering is related to spatial correlations in the quantum wavefunction, whereas purification ordering captures correlations between wavefunctions at different points in time.  Due to the asymmetric nature of the time and space directions in these models, there is no intrinsic reason to expect these two critical phenomena must coincide.  A finite separation between $p_{cp}$ and $p_{ce}$ requires an extended range of $p$ where the system purifies with a power-law decay vs.\, time that is slower than the growth of the entanglement, which would be a type of quasi-long range entanglement ordering in the system.  As a limiting example, one can consider all-to-all models, where, at $p=0$, the system can build up volume-law entanglement across all cuts in the system on $\log N$ timescales.   In this limit, any power-law purification time  could provide sufficient time to build up extensive entanglement in the system, while failing to realize a reliable encoding of quantum information.

It is also of interest to obtain a precise understanding of the critical behavior in these models.  There is considerable evidence that the measurement-induced entanglement and purification transitions in 1+1 dimensions have an emergent conformal symmetry \cite{Jian19,Bao19,Li19,Li20}.  The appropriate classification of these conformal field theories (CFTs) is not currently known.  
Although certain limiting cases are equivalent to percolation \cite{Skinner18,Bao19,Jian19}, as noted in Ref. \cite{Li19}, the observed value of $\alpha(p_c)$ for stabilizer circuits differs from that of 2D critical percolation by roughly a factor of three, suggesting that these models might be in different universality classes.  More recently, Li, Chen, Ludwig, and Fisher have found a variety of other surface scaling dimensions for the ``Clifford'' CFT that differ from percolation \cite{Li20}.     Our estimated bulk correlation length exponent $\nu=1.28(2)$, on the other hand, is very close to the $\nu=4/3$ of 2D percolation.  One possible explanation for the small difference in $\nu$ is that the Clifford critical point can be obtained by a weak perturbation of a percolation fixed point \cite{Jian19}.

We have also introduced a method to determine the order parameter exponent for the purification transition in these models \cite{Gullans19e}.  Bulk order parameter exponents in 1+1 dimensions appear consistent with percolation for both stabilizer circuits and more general models \cite{Gullans19e,Zabalo20}, but there is some evidence for a  small, but significant, difference in the surface order parameter exponent in both cases \cite{Zabalo20,Gullans19e,Li20}.   It has been argued that  these critical theories  are described by logarithmic conformal field theories with central charge zero \cite{Jian19}.  Such theories have logarithmic corrections to scaling \cite{Cardy13}, so we should be cautious in interpreting small differences in apparent critical exponents.  

Many questions remain about  the characterization of these phase diagrams and universality classes, particularly outside 1+1 dimensions or in the presence of quenched disorder.  The ubiquity of these phases in open quantum system dynamics motivate the development of a deeper understanding of the phase diagrams,  the defining properties of each phase, and the critical points.   The  close connections that arise between these problems and fault-tolerant quantum computation are also interesting to explore.  In the near term, these efforts should aid in characterizing the physics of noisy-intermediate scale quantum devices.

 \subsection{Applications to Fault-Tolerance}
 \label{sec:ftqc}

 A significant consequence of Theorem \ref{thm:code} on dynamically generated codes is that, under the stated conditions, it provides a rigorous and efficient method to sample   quantum error correcting codes that satisfy an optimal tradeoff between high code rate (obtained far below the critical point $p \ll p_{cp}$) and high error thresholds  (obtained for $p \sim p_{cp}$) either numerically or in experiment.  These codes are often highly-degenerate, require limited encoding resources, and have natural realizations in terms of stabilizer codes.  These properties make them potentially useful for fault-tolerant quantum computing (FTQC). 
 
 However, a fully fault-tolerant quantum computer or memory needs to work \emph{despite} errors in both gates and measurements, whereas in our analysis we have taken all operations to be implemented in an ideal manner that is perfectly known to the observer.  
 Far away from $p_c$, where the correlation length $\xi$ is much less than the system size, conventional scaling arguments suggest that the encoding and decoding operations away from the channel capacity limit will already converge to high-fidelity after depth  $ t/\xi^z \sim  \log N$ in the thermodynamic limit, where  $z$ is the dynamical critical exponent ($z=1$ for the $1+1$ dimensional model studied above).  Although the $\log N$ scaling implies that the encoding is unlikely to be directly achievable in a fault-tolerant manner, such low-depth codes with high code rates may be crucial in improving the performance and flexibility of intermediate-scale quantum devices \cite{Gottesman13}.  Moreover, simple extensions of the present models to include active feedback may allow the realization of fully fault-tolerant encoding schemes in this framework.
 
We have further left open the problem of efficiently characterizing the logical operator algebra of the code space density matrices $\rho_m$ or implementing the recovery operations $\mathcal{R}_{m t_{\bm{a}},t}$ in an efficient manner.    Characterizing and finding recovery operations for approximate error correction of quantum channels is an active area of research in quantum information theory (e.g., see Ref.~\cite{Junge18,Gilyen20}).  Although our proof is nonconstructive for the recovery operation, the bound that we prove on its entanglement fidelity implies that there are explicit recovery maps expressed in terms of $\rho_m$ and the $\mathcal{N}_{t,t_{\bm{a}}}$ that also satisfy the conditions of Theorem \ref{thm:code} \cite{Barnum02}.    Furthermore, in the case of stabilizer circuits exhibiting a strong purification transition, the logical operators and recovery operations can be efficiently computed using the Gottesman-Knill theorem \cite{Gottesman98,Aaronson04}.  Due to the large number of advantages of stabilizer codes for FTQC, the codes we have found emerge in stabilizer circuits are likely the most promising to explore in this context. 
 
 Our deconstruction of the recovery operation reduces the decoding problem in the mixed phase to the problem of characterizing the quantum circuits $U_{t_2 \bm{\ell}_{t_2} |m} \cdots U_{t_1 \bm{\ell}_{t_1} |m}$ arising from the mappings of the quantum channel dynamics $M_{\ell_i | \bm{\ell}_{i-1} m} \circ T_i \to U_{i \bm{\ell}_i | m}$.  Provided one can efficiently compute or approximate $U_{t \bm{\ell}_t | m}$ for each combination $(t,\bm{\ell}_t ,m )$, then the full evolution in the mixed phase can be described by a known unitary circuit acting on the system. 
In most cases, we expect that the correlations between measurement outcomes will be short range for $p<p_c$, which can be leveraged in developing efficient decoding algorithms.  
  In the case of the order parameter for strong purification transitions (given by the coherent quantum information of the system entangled with a single-reference qubit \cite{Gullans19e}),   the unitary circuit described above has a representation in terms of a product of single-qubit gates.  The simplicity of this dynamics may provide a useful starting point for investigating more general  decoding problems in the mixed phase.  These decoding problems are  reminiscent of conventional quantum error correction and suggest that studying the late time dynamics of the mixed phase  may provide useful insight into decoding problems for standard quantum error correction (and vice versa).

  We have argued that there is substantial motivation to better characterize the class of stabilizer and nonstabilizer codes that arise in the mixed phase.  Studying the associated recovery operations for the channel dynamics may also provide insight into decoding for quantum error correction in more conventional scenarios.  Due to their local encoding schemes, degeneracy, and optimal tradeoffs between high code rates and error thresholds, these codes have potential advantages for fault-tolerant quantum computation with low-overhead \cite{Gottesman13}.

\subsection{Experiments in Near-Term Quantum Devices}
\label{sec:expts}

Current quantum computing devices are far from meeting the full requirements for  scalable fault-tolerant operation \cite{Preskill18}.  The advantage of studying this class of unitary-measurement dynamics in such near-term, intermediate scale quantum  (NISQ) devices is that it combines most of the necessary ingredients to realize general purpose quantum error correction, but in a stochastic, unstructured setting.  Similar to the recent experiments on  ``quantum supremacy'' in random circuit sampling \cite{Arute19}, we suspect that realizing scalable demonstrations of measurement-induced entanglement transitions would push towards achieving what one might call ``fault-tolerant quantum supremacy.''  Such intermediate milestones are crucial in the  drive towards scalable quantum computing.  

In this vein, the purification dynamics introduced in this work serves as a useful probe of these phase transitions outside a fully fault-tolerant setting.  The exact realization of the ordered phase in $d$ spatial dimensions requires a limit where the local decoherence rate $\gamma \ll 1/L^{d+1}$ or smaller.  In contrast, assuming the local entropy production rate-density scales as $\gamma$, a  crossover between the mixed and pure phase should persist at late times with a crossover length scale  $\xi_c$ that naively scales as $ \sim \gamma^{-1/(d+1)}$ for low enough $d$.  In addition, the purification dynamics displays a direct signature of the two phases already at the level of a single-reference qubit \cite{Gullans19e}, which, together with the quantum Fisher information \cite{Bao19}, avoids the extensive overhead in both experiment and data analysis associated with measuring entanglement or entropy of large regions.    Using these local probes,  the mixed and pure phase can be efficiently studied in general models away from $p_c$ with constant depth circuits $t\gg \xi$ and a large enough number of qubits to suppress finite-size effects \cite{Gullans19e}. Such improvements in the error resilience of quantum algorithms can be the difference between applications on near-term devices and being pushed into the future realm of full fault-tolerance.   

Although not strictly necessary \cite{Gullans19e}, the most natural setting to begin studying these dynamics in near-term quantum computing devices is by implementing stabilizer circuit models such as the ones considered in this work.  The existence of highly efficient classical algorithms for these dynamics is crucial in benchmarking the performance of the experimental device in a scalable manner.  Although one might argue that the experiment is not probing new physics because its ideal dynamics can be classically simulated, this is far from the case.  The actual physical evolution of the experimental system differs dramatically from the simplistic simulations used for stabilizer circuits.  Stabilizer states are a rich class of quantum wavefunctions that can have extensive entanglement and realize many of the intrinsic subtleties associated with quantum mechanical systems that do not have natural classical analogs.  Furthermore, there are many theoretical reasons to suspect that fully fault-tolerant quantum computing will rely, at some level, on stabilizer-based quantum error correcting codes.  By studying the statistical physics of stabilizer dynamics, one is arguably gaining fundamental insight into the phase of matter associated with a fully fault-tolerant quantum computer  \cite{Aharonov00}.

\section{Conclusions}
\label{sec:conc}

We have demonstrated the existence of a class of dynamical purification phase transitions between one phase where the many-body dynamics purifies arbitrary initial states at a rate independent of the system size, and another phase where the time to purify grows exponentially in the size of the system.  We proved several general results on this class of phase transitions.  To gain deeper insight into these problems we studied specific models of stabilizer circuits that are amenable to large scale numerics, however, the general features underlying the phase transition are relevant for many physical systems undergoing high-fidelity, continuous monitoring, interspersed with entangling dynamics.  Our observation that the bipartite mutual information for completely-mixed initial states grows sublinearly in time  may aid in the development of efficient classical decoders for 1D nonstabilizer models using matrix product methods.

Furthermore, such purification transitions naturally arise in systems with long-range interactions, which are relevant to quantum computing platforms based on trapped ions \cite{Monroe14} or quantum networks \cite{Kimble08}.  The perspective on the measurement-induced entanglement transitions in terms of  purification dynamics may also aid in experimental studies, where imperfections almost inevitably drive the system towards mixed states.  

Finally, together with Ref.~\cite{Choi19}, we have shown that that the existence of the volume-law/mixed phase, where the entropy density remains nonzero, implies an extensive quantum channel capacity in the underlying open system dynamics.  In this work, we further established that the monitored dynamics itself can be used to optimally encode quantum information against the future non-unitary evolution of the system.  These results have broad implications for the study of measurement-induced phase transitions. In particular, they imply that there is a ``code space'' of operators in the system that are not measured and, thus, can preserve quantum information about the initial state.  As a result, the time-evolution in the ordered phase at late times becomes effectively reversible, unitary dynamics in the thermodynamic limit.    The dynamically generated codes are of intrinsic interest in quantum information because they saturate  fundamental bounds on the tradeoff between the density of encoded information and the threshold error rate, with potential applications to fault-tolerance.   It will be an interesting subject of future work to better characterize these codes and further explore the relations between this family of measurement-induced phase transitions and fault-tolerant quantum computation.

\begin{acknowledgments}
We thank Steve Flammia, Steve Girvin, Sarang Gopalakrishnan, Alexey Gorshkov, Matteo Ippoliti, Liang Jiang, Vedika Khemani, Stefan Krastanov, Andreas Ludwig, Chris Monroe, Adam Nahum, Pradeep Niroula, Crystal Noel, Jason Petta,  Jed Pixley, Shivaji Sondhi, Romain Vasseur, Justin Wilson, and Aidan Zabalo for helpful discussions.  We thank Yaodong Li and Matthew Fisher for helpful discussions and for pointing out a numerical error in a prior version of Fig.~\ref{fig:ent}.   We gratefully acknowledge discussions with Soonwon Choi, Yimu Bao, and Ehud Altman about the connections between purification dynamics and channel capacities.  We also thank  Soonwon Choi, Kartiek Agarwal, and Sarang Gopalakrishnan for many helpful comments on the manuscript.    Research supported in part by the DARPA DRINQS program.
\end{acknowledgments}

\appendix

\section{Dynamically Generated Codes}
\label{app:proof1}

In this Appendix, we prove Theorem \ref{thm:code} from Sec.~\ref{sec:codethm}.  This theorem shows how to dynamically generate capacity-achieving quantum error correcting codes for monitored channels with a strong purification transition.  Recall, a unitary-dephasing channel ${T}$ is one for which there exists a representation in terms of Kraus operators $\{ A_m \}$ such that, for any input state $\rho$, $T(\rho) = \sum_m A_m \rho A_m^\dag = \sum_m p_m \rho_m$ satisfies $\trace[\rho_m \rho_{m'}] \propto \delta_{m m'}$, where $\rho_m = A_m \rho A_m^\dag/p_m$ and $p_m = \trace[A_m^\dag A_m \rho]$.  This implies that there is a block diagonal representation for $T(\rho)$, which leads to an interpretation for the dynamics in terms of unitary-evolution followed by dephasing of some off-diagonal coherences.  For any unitary-dephasing channel, there is an associated  projection-valued measurement maps $M_m$  defined by the isometric projectors $P_m$ onto the support of the image of $A_m$. 
A monitored channel $\mathcal{N}_{\vec{m},t}$ indexed by measurement outcomes $\vec{m}$ and integers $t>0$  is defined by a sequence of unitary-dephasing channels $T_i$ with projection-valued measurement maps $M_{m_i}$      
\begin{align}
\mathcal{N}_{\vec{m},t}(\rho) &= M_{m_t} \circ T_t \circ \cdots \circ M_{m_1} \circ T_1 (\rho), \\
M_{m_i}(\rho)& = P_{m_i} \rho P_{m_i}^\dag \otimes \ket{m_i}\bra{m_i}.
\end{align}
The formal statement of the theorem is:
 \begin{theorem}
  Let $\mathcal{N}_{\vec{m},t} =M_{m_t} \circ T_t \circ \cdots \circ M_{m_1} \circ T_1$ be a monitored channel indexed by measurement outcomes $\vec{m}$ and integers $t>0$    with a strong purification transition. For any $p<p_c$, $\epsilon > 0 $, and any allowed pair of sequences $({\bm{a}},{\bm{b}})$ with $ t_{\bm{a}} < t_{\bm{b}}$, $a_c < a_2$,  there exists an $N_\epsilon$, such that, for any $N \ge N_\epsilon$, there is an input state $\rho_0$, a family of quantum error correcting code spaces defined by the density operators  $\rho_m \propto \mathcal{N}_{m,t_{\bm{a}}}(\rho_0)$ for $m \in \{(m_1,\cdots, m_{t_{\bm{a}}})\}$, and a family of high-fidelity recovery operations $\mathcal{R}_{ m t_{\bm{a}},t}$  for all $m$ and $ t_{\bm{a}} < t \le t_{\bm{b}} $ with an average code rate and  entanglement fidelity 
  \begin{align}
| \sum_{m} p_{m} S(\rho_{m})/N  - c(p) | &< \epsilon , \\
\sum_m p_m F_e(\mathcal{R}_{m t_{\bm{a}},t} \circ \mathcal{N}^u_{t,t_{\bm{a}}} ,\rho_m ) & \ge   1-2 \sqrt{\epsilon},
\end{align}
respectively.
Here,  $c(p)$ is the channel capacity density of the unraveled channel $\mathcal{N}_{t}^u = \sum_{\vec{m}} \mathcal{N}_{\vec{m},t}$ for $t_{\bm{a}} \le t \le t_{\bm{b}}$, $p_m = \trace[ \mathcal{N}_{m, t_{\bm{a}}}(\rho_0)]$ is the probability of obtaining the code defined by $\rho_m$ and $\mathcal{N}_{t,t_0}^u=  \sum_{\vec{m}}  M_{m_t} \circ T_t \circ \cdots \circ M_{m_{t_0+1}} \circ T_{t_0+1}$.  
\end{theorem}  
\emph{Proof.---}In the proof, we find it convenient to work mostly with the trace-preserving unravelled channels $\mathcal{N}_t^u$ and $\mathcal{N}_{t,t_0}^u$.  The proof begins from a similar line of argument as given in Sec.~\ref{sec:codethm} for many-copy quantum errror correction.  We take $N$ sufficiently large that $|Q_{t_{\bm{a}}}/N -  c(p)| < \epsilon$ and $|Q_{t_{\bm{a}}} - Q_{t_{\bm{b}}}| < \epsilon^{2}$.   Now, let $\rho_{0}$ be an input state satisfying $I_c(\rho_{0}, \mathcal{N}_{t_{\bm{b}}}) = Q_{t_{\bm{b}}}$.  Let $\ket{\psi_{RS}}$ be a purification of $\rho_0 = \sum_k \lambda_k \ket{\psi_{k0}}\bra{\psi_{k0}}$.    From the definition of monitored channels, we can purify the unraveled channel $\mathcal{N}_t^u$ as 
\begin{align}
U_{\mathcal{N}_t^u} \ket{\psi_{RS}}  \ket{0} & = \sum_{k \ell} \sqrt{p_{k \ell}} \ket{k_R}  \ket{\psi_{k \ell}}  \ket{\ell},\\ \label{eqn:icthm}
I_c(\rho_0,\mathcal{N}_t^u) &= S(\rho_S) - S(\rho_E) = \sum_{\ell} p_{\ell} S(\rho_{\ell}),\\
\rho_{\ell} &=  \sum_k p_{k{\ell}} \ket{\psi_{k{\ell}}} \bra{\psi_{k{\ell}}}/p_{\ell},
\end{align}
where $\ell \in \{ (m_1,\ldots,m_t)\}$ indexes the measurement outcomes,  $\sqrt{p_{k \ell}} \ket{\psi_{k\ell }}= \sqrt{\lambda_k} (A_\ell \ket{\psi_{k0}})\ket{\ell}$ is a tensor product of the state of the system $A_\ell \ket{\psi_{k0}}$ with the classical register state $\ket{\ell}$ used to record the measurement outcomes,  $p_{k\ell}$ is the joint probability of starting in state $\ket{\psi_{k0}}$ and finding the system in the support of $\rho_{\ell}$, and $p_{\ell} = \sum_k p_{k {\ell}}$.  

Taking $\rho_m = \mathcal{N}_{m,t_{\bm{a}}}(\rho_0)/p_m$, from Eq.~(\ref{eqn:icthm}) and our choice of input state we can establish the bound on the average code rate at time $t_{\bm{a}}$
\be
\begin{split}
|\sum_m &\frac{p_m}{N} S(\rho_{m}) - c(p)|= | I_c(\rho_0,\mathcal{N}^u_{t_{\bm{a}}})/N - c(p)|  \\
& \le \max(|Q_{t_{\bm{a}}}/N-c(p)|,|Q_{t_{\bm{b}}}/N-c(p)|) <  \epsilon,
\end{split}
\ee
where the second line follows from the fact that $I_c$ is monotonically decreasing under quantum operations (e.g., see Ref.~\cite{Schumacher96}).

  Now, consider the continued dynamics of the purification $\ket{\psi_m}=\sum_k \sqrt{p_{k m}/p_m} \ket{k_R} \ket{\psi_{k m}}$ of $\rho_{m}$ under an isometric embedding of $\mathcal{N}^u_{t,t_{\bm{a}}}$
\begin{align} \label{eqn:usetta}
U_{\mathcal{N}^u_{t,t_{\bm{a}}}} \ket{\psi_{m}} \ket{0} &=\sum_{k  \ell} \sqrt{\frac{ p_{k m \ell}}{p_m}} \ket{k_R}  \ket{\psi_{k m \ell}}  \ket{\ell},
\end{align}
where $\ell \in \{ (m_{t_{\bm{a}}+1},\cdots,m_t)\}$.
The mutual information between the  reference system and the environment for this future evolution is given by
\be
\begin{split}
I(R:E|m) &= S(\rho_{m}) - I_c(\rho_{m},\mathcal{N}^u_{t,t_{\bm{a}}})  \\
&= S(\rho_m) - \sum_{\ell} p_{\ell | m} S(\rho_{m \ell}) ,
\end{split}
\ee
where $p_{\ell | m}  = p_{m \ell}/p_m$ is the conditional Born probability of measurement record $\ell$ conditioned on the prior measurement record $m$ and $p_{m \ell} = \sum_k  p_{k m \ell}$.
The key step in the proof is that the average mutual information is then given by the decrease in coherent quantum information
\be \label{eqn:mibar}
\begin{split}
\overline{I(R:E|m)}& =\sum_m p_m I(R:E|m) \\
&= I_c(\rho_0,\mathcal{N}^u_{t_{\bm{a}}}) - I_c(\rho_0,\mathcal{N}^u_t)  < \epsilon^{2},
\end{split}
\ee
  where the bound follows because $Q_{t_{\bm{a},\bm{b}}}$ is an upper/lower bound on $I_c(\rho_0,\mathcal{N}^u_t)$ for all $t_{\bm{a}} \le t \le t_{\bm{b}}$ and we took $N$ sufficiently large that $|Q_{t_{\bm{a}}} - Q_{t_{\bm{b}}}| < \epsilon^{2}$.
  
  Since mutual information is nonnegative, we can use a Markov inequality to bound the probability that it is larger than $\epsilon$ 
  \be \label{eqn:prob}
  \mathbb{P}[I(R:E|m) > \epsilon] \le \overline{I(R:E|m)}/\epsilon < \epsilon.
  \ee
  When $I(R:E|m)< \epsilon$, then this implies that the reduced density matrix of the reference and environment is close to a product state in the sense that
  \be
  F(\rho_{RE|m}, \rho_{R|m} \otimes \rho_{E|m}) \ge 1 - 2 \sqrt{\epsilon}.
  \ee
  We now show how this bound can be used to construct a high-fidelity recovery channel following Ref.~\cite{Schumacher01}.  For completeness, we  reproduce the full argument.  First, we let $\ket{\Psi_{SR|m}} = \sum_k \lambda_{k|m} \ket{k_{R|m}} \ket{\phi_{k|m}}$ be a purification $\rho_m$ into orthonormal states.  Recall that $F(\rho,\sigma)$ is defined as the maximum square overlap over all purifications of $\rho$ and $\sigma$.  We fix a purification of $\rho_{RE|m}$ given by the state obtained from the evolution of $\ket{\Psi_{SR|m}}$ under the  isometric embedding $U_{\mathcal{N}_{t,t_{\bm{a}}}}$
  \be 
\ket{{\Psi}_{RSE|m}} = \sum_{k \ell} \sqrt{ p_{k\ell |m}} \ket{k_{R|m}} \ket{\phi_{k\ell |m }} \ket{\ell |m}.
\ee
where $\ket{\ell |m}$ is the state of the environment conditioned on observing state $\ket{m}$ at time $t_{\bm{a}}$.  Note, this is a different representation of the evolved state from Eq.~(\ref{eqn:usetta}).
By definition, there exists some purification $\ket{\hat{\Psi}_{RSE|m}}$ of $\rho_{R|m}\otimes \rho_{E|m}$ that saturates the fidelity
  \be
  F(\rho_{RE},\rho_{R}\otimes \rho_E) = |\bra{\hat{\Psi}_{RSE|m}} \Psi_{RSE|m} \rangle |^2.
  \ee 
The state $\ket{\hat{\Psi}_{RSE|m}}$ will have a Schmidt decomposition of the form
\be
\ket{\hat{\Psi}_{RSE|m}}  = \sum_{k \ell} \sqrt{\lambda_{k|m} p_{\ell |m}} \ket{k_{R|m}} \ket{\hat{\phi}_{k \ell|m}} \ket{\ell | m},
\ee
with all states in the three-fold tensor product orthonormal.
This purified state can be corrected back to the original entangled state $\ket{\Psi_{SR|m}}$ by a projective measurement onto the support of $\rho_{\ell |m}$, e.g., $P_{\ell | m} = \sum_{k} \ket{\hat{\phi}_{k \ell | m}}\bra{\hat{\phi}_{k\ell |m}}$, followed by an isometry 
\be \label{eqn:isomdag}
U_{t \ell |m}^\dag = \sum_k \ket{\phi_{k|m}} \bra{\hat{\phi}_{k\ell |m}},
\ee
which was also used in Eq.~(\ref{eqn:isometry}) of Sec.~\ref{sec:codethm}.
 Applying this recovery operation, which acts only on $S$,  to the reduced density matrix $\rho_{RS|m} = \trace_E[\ket{\Psi_{RSE|m}}\bra{\Psi_{RSE|m}}]$ results in a corrected state of the system and reference 
 \beu
 \mathcal{R}_{m t_{\bm{a}},t}(\rho_{RS|m}) = \sum_\ell  U_{t \ell |m}^\dag P_{\ell |m} \rho_{RS|m} P_{\ell |m}  U_{t \ell |m} =\rho^c_{RS|m}.
 \eeu
   Since the fidelity is monotonically increasing under quantum operations \cite{Nielsen96}, this implies the bound on the \textit{entanglement} fidelity
 \be \label{eqn:fe}
 \begin{split} 
F_e(\mathcal{R}_{mt_{\bm{a}},t} \circ &\mathcal{N}_{t,t_{\bm{a}}},\rho_m) = F(\rho^c_{RS|m}, \ket{\Psi_{RS|m}}) \\ 
& \ge F(\ket{\hat{\Psi}_{RSE|m}},\ket{\Psi_{RSE|m}} )  \\
& = F(\rho_{RE|m},\rho_{R|m} \otimes \rho_{E|m} )\\
& \ge 1 - 2 \sqrt{\epsilon}
 \end{split}
 \ee
  Using Eq.~(\ref{eqn:prob}) and Eq.~(\ref{eqn:fe}), we can bound the average entanglement fidelity as stated in the theorem.  \qed
  
  The theorem is nonconstructive for $\rho_0$ and $\mathcal{R}_{m t_{\bm{a}},t}$ (due to the assumption of channel capacity saturation of $\rho_0$ and the reliance on $\ket{\hat{\phi}_{km\ell}}$, respectively), but otherwise explicit. The theorem can be generalized to the case where $\rho_0$ does not saturate the channel capacity, but has a strong purification transition with respect to $\rho_0$ (i.e., the coherent quantum information $I_c(\rho_0,\mathcal{N}_t)$ converges to an extensive value with time-independent sub-extensive corrections).  This generalization is useful in numerical studies or experiments where capacity achieving ensembles can be difficult to find or initialize.     As discussed in Sec.~\ref{sec:disc}, the existence of this recovery operation implies that there is also an explicit recovery operation expressed in terms of $\rho_m$ and  $\mathcal{N}_{t,t_{\bm{a}}}^u$ with similar fidelities \cite{Barnum02}.

In Sec.~\ref{sec:codethm}, we introduced a stronger condition that the recovery map from $t_{\bm{b}}$ to $t_{\bm{a}}$ can be approximately decomposed into a time-local sequence of isometries acting on the system.  To avoid a large buildup in the error, we require that the average entanglement fidelity for the recovery operation at each time step satisfies $\bar{F}_e > 1 - \epsilon/(t_{\bm{b}}- t_{\bm{a}})$.  We can see how to satisfy this condition from the proof of Theorem \ref{thm:code}.  In particular, in Eq.~(\ref{eqn:mibar}) we can impose the condition that $|Q_{t_{\bm{a}}} - Q_{t_{\bm{b}}}| < \epsilon^4/16 |t_{\bm{b}}- t_{\bm{a}}|^4$, which ensures that $\bar{F}_e > 1 -  \epsilon/(t_{\bm{b}}- t_{\bm{a}})$.  As a result, this decomposition is satisfied when $|Q_{t_{\bm{a}}} - Q_{t_{\bm{b}}}|$ decays to zero faster than $1/|t_{\bm{b}}- t_{\bm{a}}|^4$ in the thermodynamic limit.  For the strong purification transition in the random Clifford model, we found in Sec.~\ref{sec:densitymat} that the difference in channel-averaged quantum capacity scales at late times $t \gg L$ as $|\bar{I}_{\max}(t_{\bm{a}})-\bar{I}_{\max}(t_{\bm{b}})| \sim \ln (t_{\bm{b}} /t_{\bm{a}}) \approx  (t_{\bm{b}}- t_{\bm{a}})/t_{\bm{a}} $.  Applying the constraint $|\bar{I}_{\max}(t_{\bm{a}})-\bar{I}_{\max}(t_{\bm{b}})| \ll  1/|t_{\bm{b}}- t_{\bm{a}}|^4$, we find that  the recovery operation can be decomposed into a time-local unitary circuit for sequences $(\bm{a},\bm{b})$  satisfying $1<a_2$,  $t_{\bm{a}} < t_{\bm{b}}$, and $(t_{\bm{b}}- t_{\bm{a}})/t_{\bm{a}}^{1/5} \to 0$ in the thermodynamic limit.

\section{Channel Capacity Bound}
\label{app:Ic}

In this Appendix, we prove that the coherent quantum information of the quantum channel $\mathcal{N}_t = \sum_{\vec{m}} K_{\vec{m}} \rho K_{\vec{m}}^\dag $  defined in Eq.~(\ref{eqn:Ntum}) is upper bounded by the average entropy of the mixed state in any unraveling of the channel
\be
I_c(\rho,\mathcal{N}_t) \le \sum_{\vec{m}} p_{\vec{m}} S(\rho_{\vec{m}}),
\ee
where $p_{\vec{m}}$ is the probability of a given quantum trajectory and $\rho_{\vec{m}} = K_{\vec{m}} \rho K_{\vec{m}}^\dag$ is the density matrix for a single trajectory.  To prove this bound we first purify the dynamics to a unitary operation on a combined reference, system, and environment, where the environment acts as a register to record the quantum trajectories
\begin{align}
\ket{\psi_{RS^{'}E^{'}}} &= \sum_{\vec{m},k}  \sqrt{\lambda_k p_{k \vec{m}}}\, \ket{k_R} \otimes \ket{\psi_{k \vec{m}}} \otimes \ket{\vec{m}}, \\
K_{\vec{m}} \ket{k_S} & = \sqrt{p_{k \vec{m}}}\, \ket{\psi_{k \vec{m}}}.
\end{align}
The coherent quantum information satisfies the identity
\begin{align*}
S(\rho_S) &- I_c(\rho_S,\mathcal{N}_t) = S(\rho_R) + S(\rho_{E^{'}}) - S(\rho_{RE^{'}})  \\
&= I(R:E') = D(\rho_{RE^{'}} | \rho_R \otimes \rho_{E^{'}}),
\end{align*}
where primes denote the reduced density matrix after applying $\mathbb{I}\otimes U_{SE}$ (note, $\rho_{R^{'}} = \rho_R$), $I(A:B)$ is the mutual information between subsystem $A$ and $B$, and $D(\rho | \sigma) = -\trace[\rho(\log \sigma -  \log \rho)]$ is the relative entropy.  The relative entropy is monotonically decreasing under the action of quantum channels due to strong subadditivity of quantum entropy \cite{Lieb73}.  Applying the dephasing channel to the environment
\be
\mathcal{E}(\rho_E) = \sum_{\vec{m}} \bra{\vec{m}} \rho_E \ket{\vec{m}} \ket{\vec{m}} \bra{\vec{m}},
\ee
results in the identity
\begin{align} \label{eqn:ineq1}
S(\rho_S) &- I_c(\rho_S,\mathcal{N}_t) = D(\rho_{RE^{'}} | \rho_R \otimes \rho_{E^{'}}) \\
&\ge D[ (\mathbb{I} \otimes \mathcal{E})(\rho_{RE^{'}}) |  \rho_R \otimes  \mathcal{E}(\rho_{E^{'}})]  \\
&= S(\rho_R) - \sum_{\vec{m}} p_{\vec{m}} S(\rho_{\vec{m}}) \label{eqn:ineq3}.
\end{align}
Since $S(\rho_S) = S(\rho_R)$ this completes the proof.

For the unraveled channel $\mathcal{N}_t^u$ (or unitary-dephasing channels more generally), there is no need to apply the dephasing channel to the environment to arrive at the identities in Eq.~(\ref{eqn:ineq1})-(\ref{eqn:ineq3}) \cite{choi}.  In this case, the additional ancilla qubits ensure that each trajectory results in a distinct, orthogonal state of the system.  As a result, we arrive at the equality $I_c(\rho,\mathcal{N}_t^u) =\sum_{\vec{m}} p_{\vec{m}} S(\rho_{\vec{m}})$. 

\section{Stabilizer Formalism}
\label{app:stab1}

In this Appendix, we provide an overview of the basic properties of stabilizer circuits. The Pauli group $\mathcal{P}_N$ acting on $N$ qubits consist of tensor products of Pauli operators 
\be
i^\ell \mathbb{I} \otimes Z \otimes X \otimes Y \otimes \cdots ,
\ee
which we abbreviate as $i^\ell Z_2 X_3 Y_4 \cdots$, with a group operation defined by the usual matrix multiplication. The group is nonabelian when accounting for the phase factors $i^\ell$ and has a total number of elements $4^{N+1}$. 
Given a quantum state $\ket{\psi}$ on a $d$-dimensional Hilbert space, there is an associated subgroup of the unitary group $U(d)$ called a stabilizer group ${\rm Stab}(\ket{\psi})$ defined as the set of unitaries $U \in U(d)$ such that $U \ket{\psi} = \ket{\psi}$.  Here, we will be interested in the much more restricted set of \textit{stabilizer states}, which are equal to the set of states that are uniquely defined by their stabilizer group when restricted to the Pauli group $S(\ket{\psi}) = {\rm Stab}(\ket{\psi}) \cap \mathcal{P}_N$.  More specifically, stabilizer states $\ket{\psi}$ are defined by abelian subgroups $S \subset \mathcal{P}_N$ such that $-1 \notin S$, $S$ has a minimal generating set of $N$ linearly independent Pauli group elements, and, for every $g\in S$,  $g \ket{\psi} = \ket{\psi}$.  Each such stabilizer group $S$ has $2^{N}$ elements

It is a convenient fact that stabilizer states are in a one-to-one correspondence with the set of states that can be generated by acting on $\ket{0}^{\otimes N}$ with Clifford group circuits \cite{Gottesman98}.  The Clifford group for $N$ qubits $\mathcal{C}_N$ is a subgroup of $U(2^N)$ that has the defining property that each $U \in \mathcal{C}_N$ maps elements of the Pauli group to other elements of the Pauli group, i.e., $U \mathcal{P}_N U^\dag = \mathcal{P}_N$.   A direct consequence of this fact is that Clifford group circuits map stabilizer states to stabilizer states.    The full Clifford group is equal to the set of unitaries generated by Hadamard  $H$, phase  $P$, and CNOT gates
\begin{align*}
H &= \frac{i}{\sqrt{2}} \left(\begin{array}{c c} 
1 & 1 \\
1 & -1
\end{array} \right),~P =\left(\begin{array}{c c} 
e^{-i \pi/4} & 0 \\
0 & e^{i \pi/4}
\end{array}\right),\\
&U_{\rm CNOT} \ket{s_1,s_2} = \ket{s_1,s_1 \oplus s_2}.
\end{align*}
where $\oplus$ is modulo 2 addition.
Another important class of operations that preserve stabilizer states are measurements of Hermitian Pauli operators.  The set of quantum circuits generated by Clifford group circuits and  measurements of Hermitian Pauli operators are known as \textit{stabilizer circuits}.  

The Gottesman-Knill theorem states that, when stabilizer circuits act on stabilizer states, an efficient classical algorithm exists to simulate their quantum dynamics \cite{Gottesman98}. The overall efficiency of this algorithm for a depth $t$ circuit with $n$ measurements scales as $O( (t+ n) N^2)$ when implemented following the approach detailed by Aaronson and Gottesman \cite{Aaronson04}.  In addition, partial traces, which map pure stabilizer states to mixed stabilizer states, and entanglement of stabilizer states can also be efficiently computed in a time $O(N^3)$ \cite{Audenaert05,Fattal04,Nahum16}.  The  mathematical structure underlying these  algorithms is an exact mapping between evolution of stabilizer states and efficient operations over $GF(2)^{2N+1}$ \cite{Gottesman96,Calderbank97}.  In this formalism, we map elements of the Pauli group 
\beu
i^\ell \mathbb{I} \otimes Z \otimes X \otimes Y \otimes \cdots \to ((0,0,1,0,0,1,1,1, \ldots) \lvert  \ell/2),
\eeu
to a binary vector through the mapping $\mathbb{I} \to (0,0)$, $Z \to (1,0)$, $X \to (0,1)$, and $Y \to (1,1)$, paired with the additional entry $\ell/2$ specifying the power of $i$ out front.  The Pauli group operation amounts to modulo 2 addition of the first $2N$ entries of this vector, while the last entry must be updated  in a way that preserves the commutation relations of the Pauli group elements.  More explicitly, representing two elements of the Pauli group as
\begin{align} \label{eqn:pi}
P_i &= ( (\bm{z}_i,  \bm{x}_i) \lvert r_i ) , ~ i \in \{1,2\}, \\
(\bm{z}_i,\bm{x}_i) & = (z_{i1},x_{i1},\ldots, z_{iN},x_{iN}),
\end{align}
then 
\begin{align}
P_1 +P_2 &= ( (\bm{z}_1 \oplus \bm{z}_2, \bm{x}_1 \oplus \bm{x}_2) \lvert r_{1} \cdot r_2 ),\\
2 r_1 \cdot r_2 &= [2 r_1 + 2 r_2 + \sum_{j} g(z_{1j},x_{1j},z_{2j},x_{2j})] \, {\rm mod}\, 4, \label{eqn:r1r2}
\end{align}
where we denote the Pauli group operation by $+$, $g(a,b,c,d)$ is a function that takes 4 bits as input and outputs the power to which $i$ is raised in the product
$\mu_{ab} \mu_{cd}= i^{g(a,b,c,d)} \mu_{a \oplus c,b \oplus d}$, where $\mu_{00} = \mathbb{I}, \mu_{01} = X, \mu_{10} = Z$, and $\mu_{11} = Y$,
\beu
\begin{split}
g(0,0,c,d)&=0,~g(1,1,c,d)=c-d,\\
g(1,0,c,d)&=d (1-2 c),~g(0,1,c,d)=c (2 d-1).
\end{split}
\eeu
This function encodes the single-site commutation relations of the Pauli group.  
Given a generating set for a subgroup of the Pauli group with $M$ elements, as well as an additional set of $2N-M$ linearly independent generators for the rest of the Pauli group, we can store all this information as a $2N \times (2N+1)$ dimensional matrix over $GF(2)$ called a \textit{tableau representation} for this subgroup.  
In Ref.~\cite{Aaronson04}, explicit algorithms are presented that take advantage of this tableau representation to evolve both pure and mixed stabilizer states in time polynomial in the number of qubits, gates, and measurements applied to the system.

\section{Mixed Stabilizer States }
\label{app:mixedevolve}

In this Appendix, we introduce some basic properties of mixed stabilizer states.  If we define an abelian subgroup  $S \subset \mathcal{P}_N$, such that $-1\notin S$, that is generated by $M < N$ elements, then the common eigenspaces of the elements of $S$ have dimension $2^{N-M}$.  A mixed stabilizer state  is a normalized projector onto the $+1$ eigenspaces of $S$:
\be
\rho = \frac{1}{2^N} \prod_i^M (\mathbb{I} + \bar{Z}_i) = \frac{1}{2^N} \sum_{g \in S} g,
\ee
where $\bar{Z}_i$ are a generating set for $S$.  Associated to the set of generators $\bar{Z}_i$  are a set of flip operators $\bar{X}_i$ that satisfy $[\bar{X}_i,\bar{X}_j ] = \delta_{ij}$ and $\bar{X}_i \bar{Z}_j = (-1)^{\delta_{ij}} \bar{Z}_j \bar{X}_i$.    
Given such a stabilizer group, we can always  extend the generating set of $M$ stabilizers $\bar{Z}_i$ and associated flip operators $\bar{X}_i$ to a complete generating set for $\mathcal{P}_N$ by finding an additional set of $2(N-M)$ operators $\{ \bar{Z}_{M+1},\bar{X}_{M+1},\ldots,\bar{Z}_N,\bar{X}_N \}$, such that whole generating set satisfies the usual Pauli algebra.   If we think of the stabilizer group $S$ associated with the mixed state  as defining a stabilizer quantum error correcting code \cite{Gottesman96}, then these additional $2(N-M)$ operators would be referred to as \textit{logical operators} in the \textit{code space}.   Here, the code space just refers to the subspace of the Hilbert space on which $\rho$ acts trivially.

The combined unitary-projective measurement dynamics applied to the completely mixed state drives the system to a mixed stabilizer state $\rho_t$ with a set of generators $\bar{Z}_i$.  If we fix a time $t$ and initialize the system at $t=0$ in an arbitrary state in any of the code spaces with generators $\{\pm \bar{Z}_i \}$, then these states will each be mapped under the dynamics to a unique state in the code space of $\rho_t$.  As a result, all initial states in the code space can be recovered using a state-independent unitary operation (fixed by the measurement record, but dependent on the gates and measurement locations) that is a  product of single-site unitaries that collectively flip the generators $\bar{Z}_i$ back to their sign in the initial state.

\section{Entanglement, Entropy, and Mutual Information of  Stabilizer States}
\label{app:ent}

In this Appendix, we describe methods to compute entropies of stabilizer states.    The density matrix for a pure stabilizer state has the explicit representation
\be
\rho = \frac{1}{2^N} \prod_i^N (\mathbb{I} + \bar{Z}_i) = \frac{1}{2^N} \sum_{g \in S} g,
\ee
where  $\{\bar{Z}_1,\ldots,\bar{Z}_N \}$ is a generating set for $S$.
For pure states, the entanglement of a region $A$ is simply the von Neumann entropy of the reduced density matrix on $A$.
From the expression for $\rho$, we can see that 
\be
\rho_A = \trace_{A^c} \rho = \frac{1}{2^{\lvert A \lvert}} \sum_{g \in S_A} g = \frac{1}{2^{\lvert A \lvert}} \prod_i^{r_A} (\mathbb{I} + \bar{Z}_{ i}),
\ee
where $S_A$ is a subgroup of $S$ with the defining property that $g$ is equal to the identity when restricted to $A^c$, i.e., $\mathbb{I}_{A^c} \otimes \trace_{A^c} g = 2^{|A^c|} g$, and $\{\bar{Z}_{1},\ldots,\bar{Z}_{r_A}\}$ is a generating set for $S_A$.    From this expression, we can see that stabilizer states have completely degenerate entanglement spectrum, such that the von Neumann entropy is given by the simple expression $S(\rho_A) = - \trace[\rho_A \log \rho_A] =\lvert A \lvert - r_A$, where $r_A$ is the number of elements in a minimal generating set for $S_A$.  Thus, to compute the entanglement it is sufficient to find the order of $S_A$.  This can be accomplished by writing a reduced tableau representation for  $\rho$ consisting of an $N \times 2N$ matrix over $GF(2)$, whose rows are the binary vectors associated with $\{\bar{Z}_1,\ldots,\bar{Z}_N \}$.  By restricting this matrix to the columns corresponding to $A^c$, we can determine $r_A$ by performing row reduction on this $N \times 2 |\bar{A}|$ matrix to put it into upper triangular form \cite{Audenaert05}.  Such a procedure will generate a linearly independent set of $\bar{r}_A$  stabilizers that that act nontrivially on $A^c$.  
The total stabilizer group $S$ is  generated by  this set of stabilizers,  together with a generating set for $S_A$, such that $N = r_A + \bar{r}_A$; thus, we arrive at the formula
\be
S(\rho_A) = \lvert A \lvert - r_A =  \bar{r}_A - |A^c|,
\ee
where $r_A$ and $\bar{r}_A$ can be computed in time $O(N^3)$ using Gaussian elimination.  

Although the above algorithm can be applied to compute the bipartite entanglement of a given partition $A \subset \{1,\ldots,N \}$, there is a natural extension of this algorithm that can be used to efficiently compute the bipartite entanglement of any contiguous region  for a fixed ordering of sites, where contiguous is defined for periodic boundary conditions with respect to the chosen ordering \cite{Nahum16,Li19}.  First, we introduce the notation of the left $\ell(g)$ and right $r(g)$ endpoints of a stabilizer $g$, which are defined as the minimal and maximal site on which $g$ acts nontrivially.  The algorithm proceeds by taking a tableau representation for the generators of a stabilizer state with respect to the chosen ordering and  performing row reduction on the entire $N \times 2N$ matrix.  This procedure effectively operates on the left end points of the stabilizers.  In the second step, row reduction is performed on the right end points, which aims to put the matrix into lower triangular form, but with an added constraint that one always eliminates  right end points by combining the stabilizer with a ``shorter'' stabilizer, where the length of a stabilizer $g$ is defined as $d(g) = r(g) - \ell(g)$.  This second round of row reduction preserves the left end points of the stabilizer generators and results in a tableau matrix in the \textit{clipped gauge}, which is defined by the condition that, for every site $x$, the stabilizer generators satisfy
\begin{align} \label{eqn:clip}
n(x) &= | \{ \bar{Z}_i : \ell(\bar{Z}_i) = x \} | +  | \{ \bar{Z}_i : r(\bar{Z}_i) = x \} | = 2,
\end{align}
with the sum rule $\sum_x n(x) = 2N$.
The constraint in Eq.~(\ref{eqn:clip}) is an immediate consequence of the above row reduction procedure.
Such a representation does not uniquely fix the stabilizer generators, but it allows for an efficient calculation of $\bar{r}_A$ for any contiguous region in terms of the positions of the left and right end points \cite{Nahum16,Li19}
\be \label{eqn:sa}
\begin{split}
S(\rho_A) = \frac{1}{2} | \{ \bar{Z}_i : \ell(\bar{Z}_i) \in A ~\&~ r(\bar{Z}_i) \in A^c \\
{\rm or}~ \ell(\bar{Z}_i) \in A^c ~\&~ r(\bar{Z}_i) \in {A} \} |
\end{split}
\ee
As a result, by performing two steps of row reduction to put the stabilizers into the clipped gauge, the entanglement can be efficiently computed for all $N(N-1)+1$ contiguous subregions by simply checking endpoint positions of the stabilizers, which reduces the overhead of the entanglement calculation by a factor of $O(N^2)$ for each subregion.  

 Entanglement of mixed states is generically difficult to compute as it requires one to distinguish classical and quantum correlations.  However, entropies and mutual informations of mixed stabilizer states can be efficiently computed using similar methods as described for pure states. The entropy formula for a given subregion $A$ has to be updated because $\bar{r}_A + r_A = M$ ($M$ is the number of independent generators for the mixed  stabilizer state)
 \be
 S(\rho_A) =  \lvert A \lvert - r_A =   \lvert A \lvert - M + \bar{r}_A ,
 \ee
 e.g., $S(\rho) = N - M$.
 Similar to pure states, we can define a clipped gauge for mixed states by performing left and then right row reduction on the first $M$ rows of the tableau representation for $\rho$.  For a given site $x$, instead of Eq.~(\ref{eqn:clip}), we  have the identity 
 \begin{align*}
n(x) &= | \{ \bar{Z}_i : \ell(\bar{Z}_i) = x \} | +  | \{ \bar{Z}_i : r(\bar{Z}_i) = x \} | \le  2, 
\end{align*}
with the sum rule $\sum_x n(x) = 2M$.

 Another difference between pure and mixed states in this representation is that it is not sufficient to know just the endpoints of the stabilizers in the clipped gauge to compute the entropy of contiguous subregions.  
 For subregions $A$ that do not wrap around site $N$, there is a formula for the entropy:
 \be
S(\rho_A) = |A| -    |\{ \bar{Z}_i : \ell(\bar{Z}_i) \in A ~\&~r(\bar{Z}_i) \in A \} |.
 \ee
 However, for contiguous regions that wrap around $N$, there is the possibility that the left and right endpoints are both in $A$, but the stabilizer has support outside of $A$.  In this case, additional linear independence tests have to be performed on $A^c$, which can take time $O(N^3)$.  As a result, for some contiguous regions, there is no advantage to working in the clipped gauge for the purposes of computing the subsystem entropy of mixed states.

 \section{Contiguous Code Length}
 \label{app:codelength}

 A mixed stabilizer state is a normalized projector onto a quantum error correcting code space.  A shorthand notation to identify the power of a code is $[N,k,d]$, which specifies that the code is defined on $N$ physical qubits, encodes $k$ logical qubits, and can correct any error that acts on up to $(d-1)/2$ physical qubits, where $d$ is the code distance.  The code distance is defined as the minimal weight (number of nonidentity sites) of all possible Pauli group elements that commute with the stabilizer group $S$, but are not contained in $S$.

 Given an $[N,k,d]$ stabilizer code $S$ with $k>0$ and a set of geometric partitions of the qubits $\bm{A} = \{ A_i : A_i \subset \{ 1, \ldots,N\} \}$, we define a  code distance with respect to $\bm{A}$ 
\be
d_{\bm A} = \min_{A_i} \{ N_{A_i}  : \exists g \in P_N,~g|_{A^c_i} = \mathbb{I},~[g,S] = 0, g\notin S\} ,
\ee
where $P_N$ is the Pauli group on $N$ qubits and $N_{A}$ is the number of elements in $A$.  If no such $A_i \in \bm{A}$ exists, then we define $d_{\bm{A}} = \max_{A_i} N_{A_i}$.  The code distance $d=d_{P}$, where $P$ is the set of all partitions of $\{ 1, \ldots,N\}$.  In general, computing the distance of an arbitrary stabilizer code is expected to be exponentially hard in $N$. 

  \begin{figure}[tb]
\begin{center}
\includegraphics[width = .45 \textwidth]{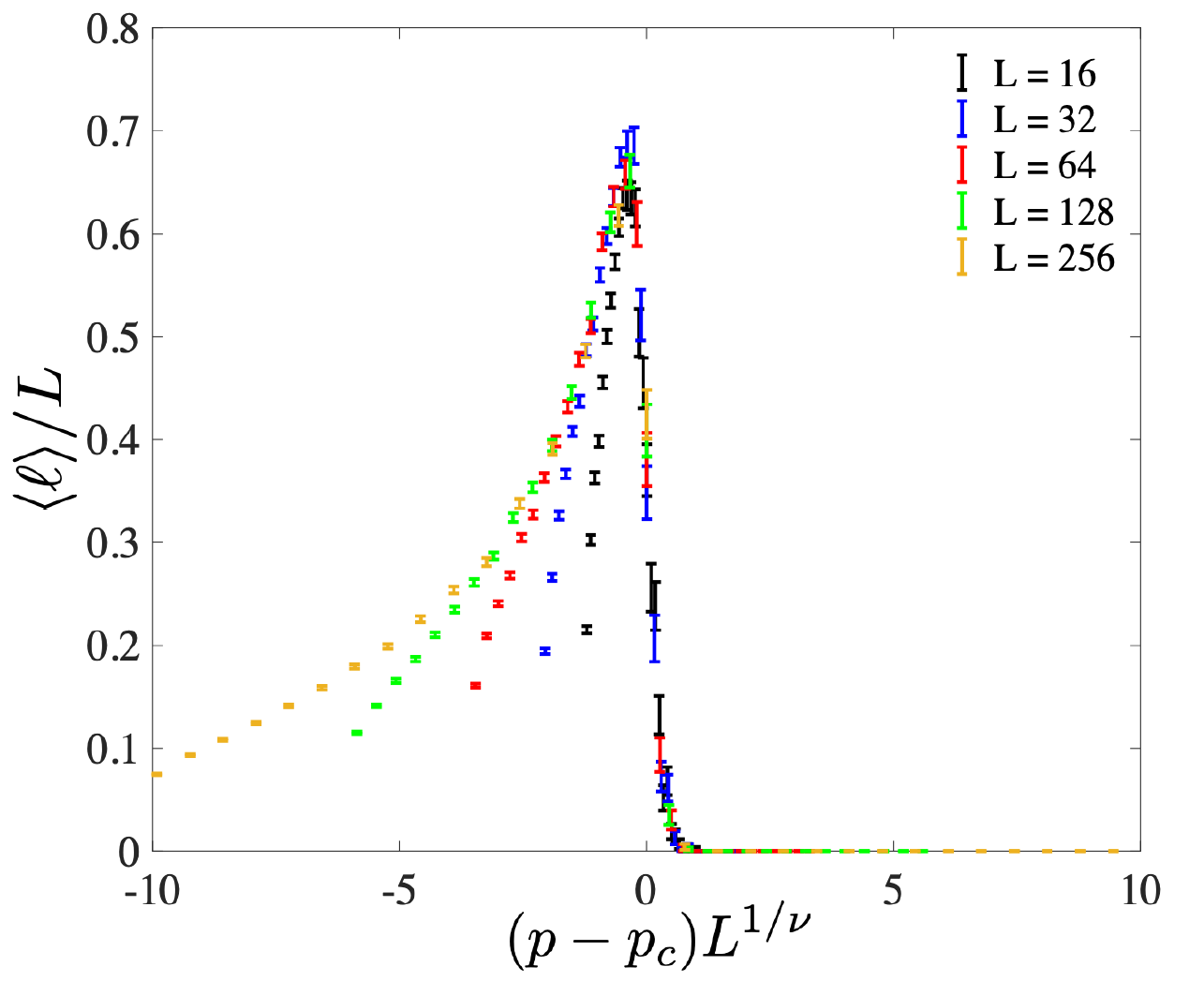}
\caption{Channel-averaged, average contiguous code length $\mean{\ell}$ for the dynamically generated codes vs. scaled measurement rate for the 1+1 dimensional stabilizer circuit model studied in Sec.~\ref{sec:crit}.  We took periodic boundary conditions with circuit depth $t=4L$.}
\label{fig:codelength}
\end{center}
\end{figure}

For the stabilizer codes considered in  Sec.~\ref{sec:crit}, which are generated by an underlying 1D random circuit, it is natural to define the contiguous code length containing site $x$ as $\ell_x \equiv d_{{\bm A}_x}$, where ${\bm A}_x$ is the set of all contiguous regions of $\{1,\ldots,L\}$ such that $x \in A_i$ for every $A_i \in \bm{A}_x$.  We distinguish periodic and open boundary conditions by whether $1$ and $L$ are considered neighbors.  The  average contiguous code length is defined as
\be
\ell = \frac{1}{L} \sum_x \ell_x.
\ee
A closely related quantity called the linear code distance $\ell_{\rm min} = \min_x \ell_x$ was introduced by Bravyi and Terhal in Ref.~\cite{Bravyi09}.
For a given partition $A$, we can determine whether there exists a logical Pauli group operator that lives on $A$ by performing Gaussian elimination on the tableau representation of a generating set for the logical operators restricted to $A^c$, which takes a time at most polynomial in $L$ \cite{Audenaert05}.  Since the set of all contiguous regions of a 1D geometry is polynomial in $L$, we can also compute the average code length for $S$ in a time polynomial in $L$.  The average contiguous code length is an upper bound on the code distance $\ell \ge \ell_{\rm min} \ge d$.

In Fig.~\ref{fig:codelength}, we show the optimal code-averaged $\mean{\ell}$  for the 1+1-dimensional model studied in Sec.~\ref{sec:crit}.  We performed the encoding step by starting from a completely mixed state and running each circuit for a time $t = 4L$.  As a convention, we define the code distance of an $[N,0]$ code (i.e., a stabilizer state) as zero.
With this convention, $\langle \ell \rangle$  equals the probability that the system is in a mixed state (and thus defines a code) times the average contiguous code length of those trajectories, which is why $\mean{\ell}$ decays to zero in the pure phase.  Deep in the mixed phase $\xi/L \ll 1$, $ \mean{\ell}$ scales subextensively as $\mean{\ell} \sim L^a$ $(a = 1/3 - 1/2)$ over this range of sizes, whereas, in the critical region of the mixed phase $\xi/L \sim 1$, $ \mean{\ell}$ seems to scale as $\mean{\ell} \sim L$.  We also find that the  average linear code distance $\mean{\ell_{\rm min}}$ (not shown) has the same scaling as the $\mean{\ell}$ throughout the mixed phase.  The extensive scaling of $\mean{\ell}$ with system size near $p_c$ in this 1+1-dimensional model adds additional evidence in support of the bound on $p_c \le 0.1893$ for the entanglement transition for qubits in 1+1 dimensions \cite{Fan20}.


\bibliographystyle{apsrev-nourl-title-PRX}
\bibliography{Purification}

\begin{thebibliography}{101}
\bibitem[{not({\natexlab{a}})}]{notelog}
\bibinfo{note}{In this work, $\log$ refers to the base-2 logarithm $\log_2$,
  while we use $\ln$ to denote the natural logarithm.}
\bibitem[{a3()}]{a3}
\bibinfo{note}{We allow $a_3(N)$ to have a nontrivial scaling with $N$,
  provided $a_3(N)/N^{a_2+1} \to 0$.}
\bibitem[{gen()}]{gendeph}
\bibinfo{note}{Note, unitary-dephasing channels are a distinct class of
  channels from the generalized dephasing channels introduced in
  \cite{Devetak05}. The latter impose the orthogonality condition on the states
  of the system instead of the environment, as in unitary-dephasing channels.}
\end{thebibliography}

\end{document}